\documentclass[10pt,conference,review]{IEEEtran}
\IEEEoverridecommandlockouts

\usepackage[T1]{fontenc}
\usepackage[utf8]{inputenc}
\usepackage{lmodern}
\usepackage{xcolor}
\definecolor{fuchsia}{HTML}{FF00FF}

\usepackage{xspace}

\usepackage{amsmath}
\usepackage{amssymb}
\usepackage{bm}

\usepackage{graphicx}
\usepackage{booktabs}
\usepackage{multirow}
\usepackage{array}
\usepackage{subcaption}
\usepackage{xcolor}
\usepackage[inline]{enumitem}
\usepackage{enumitem}
\usepackage{float}

\usepackage{pgfplots}
\pgfplotsset{compat=1.16}

\usepackage{cite}

\usepackage{listings}
\usepackage[hidelinks]{hyperref}
\usepackage[capitalize,noabbrev]{cleveref}

\usepackage{mdframed}

\usepackage{balance}

\newcommand{\linearprobing}{linear probing}

\newcommand{\val}{$\text{Acc}_{\text{Val}}$}

\newcommand{\sd}[1]{\textsubscript{$\pm$#1}}
\newcommand{\dlt}[1]{{\scriptsize$(#1)$}}
\newcommand{\meandiff}{mean-diff}
\newcommand{\bcb}{BigCodeBench}
\newcommand{\he}{HumanEval}
\newcommand{\mbpp}{MBPP+}

\newcommand{\eg}{\emph{e.g.},\xspace}

\newcommand{\indistribution}{in-distribution\xspace}

\newcommand{\outofdistribution}{out-of-distribution\xspace}

\newmdenv[
  backgroundcolor=gray!10,
  linecolor=black!50,
  linewidth=0.5pt,
  skipabove=\medskipamount,
  skipbelow=\medskipamount,
  innertopmargin=4pt,
  innerbottommargin=4pt,
  innerleftmargin=5pt,
  innerrightmargin=5pt,
  frametitle={RQ1: Key takeaway},
  frametitlebackgroundcolor=black!10
]{takeawaybox}

\begin{document}

\title{On the Robustness of LLMs' Internal Representation of Code Correctness}

\author{%
  \IEEEauthorblockN{Francisco Ribeiro, Sohaila Abdulsattar\textsuperscript{*},
                    Renata Gonzalez\textsuperscript{*},
                    Mahmoud Kassem\textsuperscript{*}, Sarah Nadi}
  \IEEEauthorblockA{New York University Abu Dhabi\\
                    Abu Dhabi, United Arab Emirates\\
                    \{francisco.ribeiro, sohaila.abdulsattar, re2230,
                    mahmoud.kassem, sarah.nadi\}@nyu.edu}
  \thanks{\textsuperscript{*}These authors contributed equally and are listed
          alphabetically.}
}

\maketitle

\begin{abstract}
Code generated by modern language models often reads naturally. Yet, it also often fails to implement what was asked. This should be no surprise, as research shows the models' own confidence signals are poorly calibrated with actual correctness. A promising way to assess correctness looks inside the model: by contrasting the hidden states of correct and incorrect programs, recent work captured an internal signal of code correctness that is able to judge candidate solutions better than the model's token-level or stated confidence, with no test execution. However, this signal was captured under one particular way, leaving open an important question: whether it reflects a robust property of the model or an artifact of that choice. We study this question systematically, varying how the signal is extracted from the model internals. Besides this, we also ask if the signal's quality is limited by the data used to extract it, by constructing program pairs that differ only in the fault that makes them incorrect. Our results show that no single configuration is best, and that isolating the fault does not help.

\end{abstract}

\begin{IEEEkeywords}
large language models, code generation, code correctness, linear probing
\end{IEEEkeywords}

\section{Introduction}
\label{sec:intro}

Writing source code is increasingly delegated to large language models (LLMs).
They complete the next line in an editor~\cite{intellicode}, generate whole functions from a description~\cite{chen2021evaluating}, and implement entire features at the repository level~\cite{jimenez2024swebench}. As they grow more fluent, their output is increasingly accepted with little scrutiny~\cite{grounded_copilot}, and a growing fraction of the code reaching real systems now originates from a model rather than a human~\cite{security_weaknesses_copilot}.

However, fluency is not the same as correctness. An LLM can produce code that reads naturally yet does not implement the requested functionality~\cite{understanding_code_generation_errors,the_counterfeit_conundrum}: it may be syntactically correct and even pass some tests, yet miss the specification, with nothing apparent in it reliably indicating whether it is correct. A good test suite can check correctness, but tests are often incomplete, missing, or themselves left to the model~\cite{oracle_problem_survey}, and running them is expensive when the LLM produces many candidates. Early signs of correctness would therefore help developers trust generated code and prioritise which candidates to review.
While the model's own confidence (i.e., the probability it assigns to the generated tokens) or its stated confidence when asked~\cite{tian2023just} are potential signals, research showed that they are not well calibrated with correctness~\cite{spiess2025calibration,llm-correctness-icse2026}.

Instead of relying on the model's outputs, another option is to look at what the LLM internally computes.
A model forms intermediate numerical representations of the input as it processes it, known as \emph{hidden states}.
These states are readily available and have been shown to encode high-level properties such as truthfulness~\cite{azaria2023internal, marks2023geometry} or latent knowledge~\cite{burns2023discovering}.
The main idea is that contrasting the hidden states of inputs that have a property against those that lack it recovers a vector along which the concept varies~\cite{zou2023representation, burns2023discovering,marks2023geometry, azaria2023internal}.
This vector, a \textit{representation} of the concept, can then be used to score new inputs by how far they point along it, indicating whether the input has the property.
We use the term \emph{\linearprobing} for this general approach of extracting a concept from hidden states.

Most \linearprobing{} work has targeted natural language (NL) concepts such as sentiment~\cite{zou2023representation} or truthfulness~\cite{azaria2023internal, marks2023geometry}.
However, recent work showed that it can also capture code correctness~\cite{llm-correctness-icse2026} where contrasting an LLM's hidden states for correct versus incorrect implementations of the same task isolates a \emph{code correctness representation}. 
\Cref{fig:motivating} illustrates this idea, showing a model-generated implementation
and the benchmark's canonical solution for one \he\ task: returning the sum of the
odd values that sit at even positions. The model's solution steps through the odd positions
instead of the even ones. 
The authors' captured code correctness representation ranked the correct program above the incorrect one without needing test execution, while the model's own output probability and stated confidence did not.
However, the highest obtained accuracy ranged from 41--63\% across models and benchmarks, suggesting that the method of capturing the representation may not generalize well.
Furthermore, for a fixed model/benchmark combination, accuracy varied significantly by which data subset captured and validated the representation versus tested it.

\begin{figure}[t]
\centering
\begin{subfigure}[t]{\columnwidth}
\begin{lstlisting}[language=Python, basicstyle=\fontsize{6.5}{7.5}\selectfont\ttfamily]
def solution(lst):
  sum_of_odd_elements = 0
  for i in range(1, len(lst), 2):
      if lst[i] % 2 != 0:
          sum_of_odd_elements += lst[i]
  return sum_of_odd_elements
\end{lstlisting}
\vspace{-1em}
\caption{\scriptsize Incorrect: LLM-generated attempt that iterates over \emph{odd} positions.}
\label{fig:motivating-incorrect}
\end{subfigure}

\begin{subfigure}[t]{\columnwidth}
\begin{lstlisting}[language=Python, basicstyle=\fontsize{6.5}{7.5}\selectfont\ttfamily, numbers=left]
def solution(lst):
  return sum([x for idx, x in enumerate(lst) if idx%2==0 and x%2==1])
\end{lstlisting}
\vspace{-1em}
\caption{\scriptsize Correct: canonical solution that iterates over \emph{even} positions.}
\label{fig:motivating-correct}
\end{subfigure}
\caption{Two implementations of \he/121 --- \emph{return the sum
of the odd values at even positions}.\vspace{-0.5cm}}
\label{fig:motivating}
\end{figure}

Accordingly, the current results in the literature are promising but leave open questions about the representation's quality and its robustness to data variation.
Our goal is to systematically investigate the factors that influence the quality and robustness of code correctness representations.
We make two key observations that motivate our study design.
First, there are multiple methods to construct a concept representation in the literature~\cite{zou2023representation, im2026unifiedunderstandingevaluationsteering}.
We hypothesize that the choice of method may affect the quality of the captured representation, and that perhaps some methods may generalize better than others across models and benchmarks.
Second, the contrast between correct and incorrect programs may carry incidental differences that are not relevant to correctness, such as variable names or implementation style.
Such captured incidental differences may explain why the correctness representation becomes sensitive to data variation.
For example, in the example in Figure~\ref{fig:motivating}, the main fault is on Line 4, where the model iterates over odd positions instead of even ones.
However, the two programs also differ in other ways: the correct program uses a comprehension with \texttt{sum} while the incorrect one uses a loop with an accumulator, they use different variable names, and they check for parity differently (\texttt{\%2!=0} vs. \texttt{\%2==1}).

In this paper, we systematically investigate the design choices behind capturing a code correctness representation to assess their effect.
We also investigate how data variation affects the quality of the captured correctness signal.
More specifically, we answer the following research questions (RQs):

\begin{enumerate}[leftmargin=*, align=left, label={\textbf{RQ\arabic*}}]
\item How do the key design choices involved in capturing a code correctness direction (the direction-construction method, prompt framing, and hidden-state location) affect its quality and robustness?

\item Does controlling for differences between correct/incorrect programs via mutations and refactorings improve the quality of the correctness direction?
\end{enumerate}

Our evaluation covers \emph{in-distribution} settings, where the direction is fit and tested on disjoint subsets of the same benchmark (\he~\cite{chen2021evaluating} and \bcb~\cite{zhuo2024bigcodebench}), and \emph{out-of-distribution} settings, where it is fit on \mbpp~\cite{austin2021program,liu2023your} and tested on either benchmark, across four instruction-tuned LLMs.

 Our results show that no single configuration recovers the direction best. Only the construction method generalises, while the best prompt framing, read-out location, and model change from one benchmark to the other. Controlling the fitting pairs to isolate the bug-causing change does not improve the direction either --- on \bcb\ a direction fit on such pairs separates them with high accuracy in isolation, yet does not transfer to ranking the benchmark's own candidate implementations.

Overall, the contributions of this paper are threefold:
\begin{enumerate*}
  \item a systematic study of how the \emph{method} of constructing
    the direction, the \emph{framing} of the prompt, and the \emph{hidden-state location} read from the model affect the accuracy of the captured code correctness direction across four LLMs and two benchmarks.
  \item a data generation pipeline that isolates bug-causing
    changes via controlled mutations and behavior-preserving refactorings to test if directions built from
    such data, rather than from model-generated failing attempts, carry a higher quality correctness signal.
  \item a publicly available replication package containing the data and code to reproduce the study:
\end{enumerate*}\vspace{-0.1cm}
\begin{center}
    \underline{\textit{\url{https://figshare.com/s/8da5228fa8cc8fc668f8}}}\vspace{-0.1cm}
\end{center}

\section{Background}
\label{sec:background}
This section presents foundational aspects to understand how LLMs represent data internally, how to extract a direction corresponding to a target concept, and how prior work applied this to code correctness.

\subsection{Hidden states locations}
\label{subsec:locations}
A decoder-only transformer language model maps a token sequence to hidden states
through a stack of $L$ layers~\cite{vaswani2017attention}. For an input $x$, it
produces a hidden state $h^{(l)}_i(x) \in \mathbb{R}^{d}$ at every layer
$l \in \{0,1,\dots,L\}$ and every token position $i$, where $l=0$ is the
embedding layer and $d$ is the hidden size. 
These states encode high-level semantic and syntactic properties recoverable with simple probing methods~\cite{zou2023representation}. As a hidden state exists for every token at every layer, it requires deciding which token to read the hidden state from. Two choices are common: reading the \emph{last token}, a single hidden state that has attended to the entire preceding stimulus, or the \emph{response average}, the mean of the hidden states over the response tokens~\cite{zou2023representation}.

\subsection{Construction methods}
\label{subsec:methods}

Linear probing captures a direction in hidden-state space corresponding to a
target concept by contrasting representations of inputs that differ in
it~\cite{zou2023representation}. Starting from paired \emph{positive} and
\emph{negative} stimuli, it takes a hidden-state representation for each at a chosen
location (\Cref{subsec:locations}) and turns the differences into one concept
direction $v^{(l)}$ per layer.
Truthfulness, sentiment, and other abstract concepts are recoverable in this linear form
and can be used to monitor model behaviour~\cite{burns2023discovering,
li2023inference, marks2023geometry, azaria2023internal}.

 Two
prominent methods turn contrasting pairs into a direction:
\begin{enumerate*}[label=\textbf{(\alph*)}]
\item \textbf{Principal component analysis (PCA):} the per-pair representation
differences at layer $l$ are collected and mean-centered, and $v^{(l)}$ is taken as
their first principal component (i.e., the direction of largest variance among the
contrasts)~\cite{abdi2010principal}. Its orientation is arbitrary, so the sign is
resolved by a per-layer majority rule over labeled examples, so that higher
projections correspond to inputs that exhibit the concept~\cite{zou2023representation}.

\item \textbf{Mean-difference (\meandiff):} the direction is the difference between the
positive and negative class means at layer $l$,
\end{enumerate*}
\begin{equation}
  v^{(l)} \;=\; \bar{a}^{(l)}_{+} - \bar{a}^{(l)}_{-},
  \label{eq:meandiff}
\end{equation}
where $\bar{a}^{(l)}_{+}$ and $\bar{a}^{(l)}_{-}$ are the class means. Its
orientation is fixed, so no separate sign step is needed. Despite its simplicity,
\meandiff{} has been reported to often recover directions better than PCA~\cite{im2026unifiedunderstandingevaluationsteering}.
Regardless of how a direction is captured, a new input is scored by projecting its representation onto the
direction: the further it points along it, the more it exhibits the concept.

\subsection{Framings}
\label{subsec:framings}

We use the term \emph{framing} for how the input is wrapped into a prompt.
Framing shapes which aspects of the model's internal state the contrast picks
\Cref{fig:framings} shows three common framings in this line of work:
\begin{enumerate*}[label=\textbf{(\alph*)}]
\item \textbf{Neutral:} The model is shown only the input, with no reference to the
concept~\cite{llm-correctness-icse2026}, so any signal must arise from the model's
own unguided processing. This is the most conservative framing.

\item \textbf{Concept:} The stimulus adds a meta-instruction directing
the model to consider how much of the concept is present, eliciting its declarative
knowledge about that concept~\cite{zou2023representation}.

\item \textbf{Function:} The stimulus instructs the model to produce behaviour that exhibits the concept or its opposite (e.g. to write a correct or an incorrect implementation), contrasting the instruction-conditioned states~\cite{zou2023representation}.
\end{enumerate*}

\subsection{Using \linearprobing{} to capture code correctness}
\label{subsec:priorwork}

While most \linearprobing{} work targets abstract NL concepts, e.g. truthfulness~\cite{azaria2023internal, marks2023geometry}, Ribeiro et al.~\cite{llm-correctness-icse2026} adapted it to capture code
correctness, the same concept our work studies.
Their contrasting pairs join a task's canonical solution with a \emph{plausible failing attempt}: an LLM-generated implementation of the task that fails its test suite. Because such an attempt is generated independently of the canonical solution, it can differ from it in any way --- algorithm, code structure, variable naming --- and not only in what makes it fail.
The authors reported two findings. First, a code correctness
direction can be \emph{captured} and used to distinguish correct from incorrect
programs more accurately than the model's own confidence, whether read
\emph{intrinsically} from token probabilities~\cite{kadavath2022language} or
\emph{reflectively} from a verbalised judgement~\cite{tian2023just}. Second, it is
\emph{useful for ranking}: ordering candidate implementations by their correctness score raises the chance a correct one is selected, matching or exceeding a specialized ranker~\cite{rankef}.
However, their results came from a \emph{single} choice on each axis: the
direction was built with PCA, read from the last token, and fit under a neutral
framing but evaluated under a concept one. Its accuracy also varied with the data split, so the signal may be sensitive to how it is captured, both in terms of method and data.

\section{Experimental Setup}
\label{sec:experimental-setup}

This section describes our experimental setup. We first cover the aspects
shared across both RQs
(Sections~\ref{sec:setup-protocol}--\ref{sec:setup-stimuli}), and then
settings specific to each RQ (Sections~\ref{sec:rq1-data}
and~\ref{sec:rq2-data}).

\subsection{Fitting, Validation, and Testing Protocol}
\label{sec:setup-protocol}
We split tasks into disjoint fitting, validation, and test partitions. The
per-layer direction is captured on the fitting partition. Since the most
informative layer is not known in advance, the validation partition selects a
single layer $l^\star$ (the one whose direction best separates correct from
incorrect programs under the validation framing
(Section~\ref{sec:setup-design})) and we report accuracy on the held-out test
partition at $l^\star$ under the matching test framing.

A program is scored by projecting its representation at the chosen read-out
location (Section~\ref{sec:setup-design}) onto
the layer's direction,
\begin{equation}
  s^{(l)}(x) \;=\; \sigma^{(l)}\,
    \frac{\big\langle h^{(l)}(x),\, v^{(l)} \big\rangle}{\lVert v^{(l)} \rVert},
  \label{eq:score}
\end{equation}
where $\sigma^{(l)} \in \{+1,-1\}$ is the per-layer sign: the majority-rule value
for PCA (Section~\ref{subsec:methods}), and $+1$ for \meandiff{}, whose orientation is already fixed. A higher
absolute score is intended to indicate a more likely correct program.

We use two metrics. \textbf{Pairwise accuracy} is the fraction of same-task
correct/incorrect pairs in which the correct program scores higher (a 1-of-2
choice); during validation, we select $l^\star$ as the layer with the highest
pairwise accuracy on the validation partition, breaking ties toward the deepest
layer. \textbf{Selection accuracy (\val)} is the test-time metric: among four
candidate completions for a task, the task succeeds when the correct one is
selected (a 1-of-4 choice). We report it at test time because it matches the
intended use and the setting of prior work~\cite{llm-correctness-icse2026,zou2023representation}. We report selection accuracy under two layer choices: \val{} is selection
accuracy at $l^\star$, the layer chosen on the validation partition without test
labels, and is our primary reported number; \textbf{best-layer accuracy} is
selection accuracy at the layer maximising accuracy on the test partition, an oracle
upper bound isolating direction quality from layer-selection cost.

Prior work established that the direction can outperform the model's own confidence~\cite{llm-correctness-icse2026}, so we do not repeat that comparison. Instead, we focus on how construction and data affect the direction itself.

\subsection{Design Space}
\label{sec:setup-design}

A configuration pairs a construction method (PCA or mean-diff), a framing
(neutral, concept, or function), and a hidden-state location (last
token or response average). The construction method asks which notion of separation aligns with correctness, and each method embodies a different trade-off. PCA keeps only the dominant axis of variation among the pair differences: if correctness drives most of the variation between correct and incorrect programs, PCA isolates it and discards the rest as noise, but if off-target properties such as length or formatting dominate, the correctness signal lies along one of the discarded axes and is lost. Mean-diff discards nothing: the difference of the class means retains, in one mixed direction, every property that systematically differs between correct and incorrect programs --- correctness along with any incidental difference that accompanies it. Thus, the trade-off is between losing the signal when it is not dominant (PCA) and diluting it with whatever else separates the correct and incorrect classes (mean-diff). Because neither risk can be ruled out in advance, we evaluate both. The framing asks how explicitly, if at all, the LLM is prompted for
correctness. The read-out location asks whether a single fully-informed position
suffices (last token) or pooling across the response helps (response average).

A framing mode specifies two framings: one used when fitting the direction, and one used when validating and testing it (Section~\ref{sec:setup-protocol}); the latter two stages (validating and testing) always share the same framing, so the layer is selected under the framing the direction is scored under. Rather than crossing every fit framing with every evaluation framing, we
use six modes, written \emph{fit}$\to$\emph{eval}: \texttt{N$\to$N},
\texttt{N$\to$C}, \texttt{C$\to$N}, \texttt{C$\to$C}, \texttt{F$\to$N}, and
\texttt{F$\to$C}, where \texttt{N}, \texttt{C}, and \texttt{F} denote the
neutral, concept, and function framings. The function framing is restricted to
fitting, since it pairs a positive instruction with the correct program and a
negative one with the incorrect program, so the instruction itself reveals the
label and would leak it at validation or test time; it therefore never appears
as an evaluation framing.

RQ1 varies these three axes with the data fixed to the correct/incorrect
pairs of prior work~\cite{llm-correctness-icse2026}. RQ2 fixes the
configuration and varies the data, using pairs we construct instead
(Sections~\ref{sec:rq1-data} and~\ref{sec:rq2-data}).

\definecolor{stimyellow}{HTML}{FCEFA0}
\definecolor{stimblue}{HTML}{BFE1F6}
\definecolor{stimpink}{HTML}{F9C6E0}
\newcommand{\codefont}{\fontfamily{zi4}\selectfont}
\newcommand{\stimseg}[2]{\colorbox{#1}{\parbox[t]{\dimexpr\linewidth-2\fboxsep\relax}{\fontsize{6}{6.4}\selectfont\codefont\raggedright #2}}}
\newcommand{\stimcore}{\stimseg{stimyellow}{\textbf{Task:} <task>\\ \textbf{Code:}\\ \textasciigrave\textasciigrave\textasciigrave python\\ <code>\\ \textasciigrave\textasciigrave\textasciigrave}}

\begin{figure}[t]
\centering
\setlength{\tabcolsep}{2pt}
\setlength{\fboxsep}{1.5pt}
\renewcommand{\arraystretch}{1.0}
\begin{tabular}{|c|c|c|}
\hline
{\fontsize{7}{8}\selectfont\bfseries Neutral} & {\fontsize{7}{8}\selectfont\bfseries Concept} & {\fontsize{7}{8}\selectfont\bfseries Function} \\ \hline
\begin{minipage}[t]{0.30\columnwidth}\vspace{-4pt}
\stimcore
\vspace{1pt}\end{minipage} &
\begin{minipage}[t]{0.30\columnwidth}\vspace{-4pt}
\stimseg{stimblue}{Consider the amount of correctness in the following code.}\\[1pt]
\stimcore\\[1pt]
\stimseg{stimpink}{The amount of correctness in the code is}
\vspace{1pt}\end{minipage} &
\begin{minipage}[t]{0.30\columnwidth}\vspace{-4pt}
\stimseg{stimblue}{Write a correct/incorrect implementation for the following task.}\\[1pt]
\stimcore
\vspace{1pt}\end{minipage} \\ \hline
\end{tabular}
\caption{The three framings' core is a task and candidate program
(yellow). \emph{Neutral} is the core only. \emph{Concept} adds a meta-instruction
(blue) and an answer cue (pink). \emph{Function} prepends a persona instruction (blue).}
\label{fig:framings}
\end{figure}

\subsection{Models and Benchmarks}
\label{sec:setup-models}

We use the same four instruction-tuned LLMs as prior work~\cite{llm-correctness-icse2026}, with their short names in bold: \textbf{Mistral}-7B-Instruct-v0.3, \textbf{Qwen}-2.5-Coder-7B-Instruct, \textbf{OpenCoder}-8B-Instruct, and \textbf{CodeLlama}-7B-Instruct. All are open-weights (required to read hidden states) and lie in the 7--8B range, so scale does not confound cross-model comparison. Mistral is the general-purpose model from the original representation engineering work~\cite{zou2023representation}, and the other three are code-specialized. We evaluate on the same two benchmarks: \he~\cite{chen2021evaluating}, with 164 self-contained problems over core Python and foundational algorithms, and \bcb~\cite{zhuo2024bigcodebench}, with 1,140 tasks that call external libraries such as \texttt{numpy} and \texttt{pandas}. Each task provides a correct solution and an executable test suite, which we use as a proxy for correctness. As a methodological check, we reproduce the single configuration of prior work~\cite{llm-correctness-icse2026}, a PCA direction read from the last-token hidden state fit under a neutral framing and evaluated under a concept one, and confirm reproduction of its published accuracies.

\subsection{Stimulus Construction}
\label{sec:setup-stimuli}

A stimulus is the prompt from which we collect hidden states: it presents the task and a candidate program to the LLM under one of the three framings of Section~\ref{subsec:framings}; which framing is used is set by the configuration (Section~\ref{sec:setup-design}). \Cref{fig:framings} shows the three framings on the same task and program.

The replication package of prior work~\cite{llm-correctness-icse2026} assembled the stimulus as a plain concatenated string. Throughout this paper we instead build it structurally, e.g.\ using \texttt{chat\_template} for the neutral framing.

\subsection{RQ1: Data Settings}
\label{sec:rq1-data}
We borrow a significant amount of the data setup used by Ribeiro et al.~\cite{llm-correctness-icse2026} to isolate the effect of the design choices. Each fitting pair joins the task's canonical solution with a plausible failing attempt (Section~\ref{subsec:priorwork}). The data is split into fitting, validation, and test partitions in two ways:

\textbf{In-distribution:} the direction is fit, validated, and tested on folds of the same dataset under ten-fold cross-validation.

\textbf{Out-of-distribution:} RQ1 additionally fits on two out-of-distribution sources, MBPP+~\cite{austin2021program,liu2023your} and a fully LLM-generated naive task set, under nested cross-validation: fitting and validation on four inner folds and ten outer folds. Testing always uses ten outer folds of the target dataset (\he\ or \bcb).

Therefore, the full RQ1's configuration space is $2 \times 2 \times 3 \times
2 \times 3 \times 2 \times 4 = 576$ configurations, corresponding to
method, read-out location, fit framing, validation/test framing,
fit source, test dataset, and model, respectively.

\subsection{RQ2: Data Settings}
\label{sec:rq2-data}

RQ2 keeps the configuration fixed and changes the data the direction is fit and evaluated on. Starting from the \he\ and \bcb\ tasks, we construct correct/incorrect pairs in four steps:

\textbf{Step 1: Mutant Generation.}
We introduce a bug into a correct solution with a rule-based mutant generator that applies one of 20 operators, each a small and targeted edit~(Table~\ref{tab:mutation-taxonomy}).
\begin{table}[t]
\centering
\caption{The 20 mutation operators used to build the controlled pairs.}
\label{tab:mutation-taxonomy}
\scriptsize
\setlength{\tabcolsep}{4pt}
\begin{tabular}{@{}lll@{}}
\toprule
\textbf{Code} & \textbf{Operator} & \textbf{Example (orig.\ $\rightarrow$ mut.)} \\
\midrule
\multicolumn{3}{@{}l}{\textit{Arithmetic \& Relational}}\\
\texttt{AOR} & Arithmetic Op.\ Replace & \texttt{a + b} $\rightarrow$ \texttt{a - b} \\
\texttt{AOD} & Arithmetic Op.\ Delete & \texttt{a + b} $\rightarrow$ \texttt{a} \\
\texttt{AOI} & Arithmetic Op.\ Insert & \texttt{a} $\rightarrow$ \texttt{a + 1} \\
\texttt{ROR} & Relational Op.\ Replace & \texttt{a == b} $\rightarrow$ \texttt{a != b} \\
\texttt{BOR} & Bitwise Op.\ Replace & \texttt{a | b} $\rightarrow$ \texttt{a \& b} \\
\texttt{NEG} & Negation & \texttt{x} $\rightarrow$ \texttt{-x} \\
\addlinespace
\multicolumn{3}{@{}l}{\textit{Variables \& Constants}}\\
\texttt{CR}  & Constant Replace & \texttt{x = 5} $\rightarrow$ \texttt{x = 6} \\
\texttt{VCR} & Variable to Constant & \texttt{return x} $\rightarrow$ \texttt{return 0} \\
\texttt{CVR} & Constant to Variable & \texttt{return 0} $\rightarrow$ \texttt{return x} \\
\texttt{VVR} & Variable to Variable & \texttt{a = b} $\rightarrow$ \texttt{a = c} \\
\texttt{CRN} & Constructor to None & \texttt{Node()} $\rightarrow$ \texttt{None} \\
\addlinespace
\multicolumn{3}{@{}l}{\textit{Control Flow \& Logic}}\\
\texttt{SD}  & Statement Delete & \texttt{x = 1; y = 2} $\rightarrow$ \texttt{y = 2} \\
\texttt{COR} & Conditional Op.\ Replace & \texttt{a and b} $\rightarrow$ \texttt{a or b} \\
\texttt{COD} & Conditional Op.\ Delete & \texttt{a and b} $\rightarrow$ \texttt{a} \\
\texttt{UOI} & Unary Op.\ Insert & \texttt{if x:} $\rightarrow$ \texttt{if not x:} \\
\texttt{UOD} & Unary Op.\ Delete & \texttt{not x} $\rightarrow$ \texttt{x} \\
\texttt{RC}  & Remove Conditional & \texttt{if a:} $\rightarrow$ \texttt{if True:} \\
\texttt{RVM} & Return Value Mutate & \texttt{return x} $\rightarrow$ \texttt{return None} \\
\texttt{TR}  & True Return & \texttt{return x} $\rightarrow$ \texttt{return True} \\
\texttt{FR}  & False Return & \texttt{return x} $\rightarrow$ \texttt{return False} \\
\bottomrule
\end{tabular}
\end{table}
The generator parses the program into an Abstract Syntax Tree (AST) with type inference via \texttt{astroid}~\cite{astroid}, keeping mutations syntactically valid and applied only to semantically compatible types. For example, it blocks arithmetic mutations (swapping \texttt{+} for \texttt{-}) on string or list operands, which would fail with a trivial \texttt{TypeError} rather than a plausible logical bug; since Step~2 keeps any mutant that fails the test suite, such trivially broken edits would otherwise pass that filter alongside genuine logical faults. This step mutated 1,135 \bcb\ tasks (20 operators applied; statement deletion (SD) most abundant at ${\sim}31\%$) and 143 \he\ tasks (19 operators applied; constant replacement (CR) most abundant at ${\sim}26.8\%$), dropping only tasks with no applicable mutant.

\textbf{Step 2: Mutant Validation.}
We execute every mutant against the benchmark's native test suite and keep only those that explicitly fail --- \emph{killed} mutants, in mutation-testing terms~\cite{just2014mutants} --- discarding semantic equivalents that pass. This yielded 27,928 killed mutants for \bcb\ and 4,566 for \he.

\textbf{Step 3: Confounding.}
A model may exploit surface shortcuts, e.g. keying on variable naming or code length as the ``bug'' signal rather than the semantics. To counter this, we add confounds: structurally refactored versions of a mutant that fail the same way. We prompt \texttt{Qwen2.5-Coder-32B-Instruct}~\cite{hui2024qwen2} to rewrite each killed mutant with behaviour-preserving edits, giving a distinct version of the same error.

\textbf{Step 4: Confound Validation.}
The rewrite may introduce a different error or accidentally fix the bug, so we re-run each confound and keep it only if it \textbf{fails with the same failure fingerprint} as its source mutant, discarding any that pass or fail differently. This leaves 9,585 validated confounds for \bcb\ (593 tasks) and 1,262 for \he\ (66 tasks).

\textbf{Data Splitting.}
Mirroring RQ1, we split tasks into 10/10/80 fitting/validation/test folds under ten-fold cross-validation. From the validated pairs, we form two variants of \emph{controlled pairs}: \textbf{M-Only}, pairing each reference solution with its mutants, and \textbf{MC}, which also adds the confounds. Each variant is evaluated in isolation --- fit, validated, and tested on held-out splits of the same pairs --- and in transfer --- fit and validated on the pairs but tested on the benchmark's own 1-of-4 selection task from RQ1, the same test its in-distribution directions face. Since a task yields many mutants and confounds, we sample the fitting and validation pairs in two ways: \emph{weighted} keeps one mutant per task (probability proportional to its operator's frequency) and, for MC, one uniformly random confound; \emph{paired} uses every mutant and confound, giving more but uneven pairs. Weighted is the default for the fitting and validation pairs, and the test split of controlled pairs is always sampled weighted; paired is a robustness check (\Cref{sec:rq2}).

\section{RQ1: How do the key design choices involved in capturing a code correctness direction affect its quality and robustness?}
\label{sec:rq1}

Tables \ref{tab:rq1-bcb} and \ref{tab:rq1-he} show the full sweep of results: every valid combination of construction method, read-out location, and
framing (\Cref{sec:setup-design}) across four models. Each cell is \val,
the validation-selected accuracy (\Cref{sec:setup-protocol}); the first row of each table is
the configuration used by prior work~\cite{llm-correctness-icse2026}. The tables show that no single configuration is best across both benchmarks or across LLMs. In-distribution, a single configuration is best across all four models on \bcb\ (\meandiff, response average, neutral fitting and testing framing), but on \he\ the best configuration differs by model. We therefore ask how much each choice, on its own or combined with others, changes \val.

We measure this with an analysis of variance (ANOVA), reported in \Cref{tab:rq1-nav}.
Each choice is a \emph{factor} --- construction method, read-out location,
fitting framing, testing framing, benchmark, and model. ANOVA gives every factor, and every combination of factors, a share of
the variance in \val: its $\eta^2$. A single factor's share is its \emph{main effect};
a combination's share is an \emph{interaction}, meaning the factors act together, so
neither can be set without regard to the other. Each factor and each interaction is one \emph{term} of the decomposition; the six
factors and all possible interactions give $63$ terms, whose $\eta^2$ sum to $100\%$. \Cref{tab:rq1-nav} lists the
largest terms.%

\begin{table*}[t]
  \centering
  \caption{RQ1 configuration sweep on \bcb. Per-model \val\ (mean
  and std-dev (subscript) over folds) for every construction method
  (\textsc{pca}/\meandiff\ (\textsc{md})), read-out location (last token (\textsc{last})/response average
  (\textsc{avg})), fitting framing, and validation=testing framing
  (\textsc{neut}/\textsc{conc}/\textsc{func}). First row $\approx$ prior
  work~\cite{llm-correctness-icse2026}; directions fit in- or out-of-distribution
  (\mbpp/synthetic); per-column maxima in \textbf{bold}. All numbers are percentages.}
  \label{tab:rq1-bcb}
  \scriptsize
  \setlength{\tabcolsep}{2pt}
  \resizebox{\textwidth}{!}{%
  \begin{tabular}{@{}llll *{12}{c}@{}}
    \toprule
     & & & & \multicolumn{4}{c}{in-distribution} & \multicolumn{4}{c}{OOD: \mbpp} & \multicolumn{4}{c}{OOD: synthetic} \\
    \cmidrule(lr){5-8}\cmidrule(lr){9-12}\cmidrule(lr){13-16}
    method & loc & fit & val/test & Mistral & Qwen & CodeLlama & OpenCoder & Mistral & Qwen & CodeLlama & OpenCoder & Mistral & Qwen & CodeLlama & OpenCoder \\
    \midrule
    \multicolumn{16}{@{}l}{\textit{Configuration of prior work~\cite{llm-correctness-icse2026}}}\\
    \textsc{pca} & \textsc{last} & \textsc{neut} & \textsc{conc} & 42.7\sd{4.5} & 40.8\sd{5.8} & 40.5\sd{8.5} & 39.6\sd{5.1} & 32.2\sd{3.6} & 20.5\sd{7.6} & 30.1\sd{6.8} & 28.4\sd{4.3} & 27.6\sd{12.2} & 10.3\sd{1.9} & 22.8\sd{4.1} & \textbf{36.0}\sd{5.8} \\
    \midrule
    \multicolumn{16}{@{}l}{\textit{Full configuration sweep}}\\
    \textsc{pca} & \textsc{last} & \textsc{neut} & \textsc{neut} & 38.8\sd{4.5} & 34.4\sd{3.3} & 38.3\sd{3.6} & 39.5\sd{5.1} & 23.5\sd{1.9} & 22.7\sd{1.3} & 20.3\sd{1.4} & 19.1\sd{1.1} & 32.0\sd{1.8} & 22.7\sd{2.8} & 26.3\sd{3.6} & 24.1\sd{3.8} \\
    \textsc{pca} & \textsc{last} & \textsc{conc} & \textsc{neut} & 37.0\sd{4.1} & 47.2\sd{5.4} & 36.5\sd{3.3} & 47.7\sd{3.6} & 24.7\sd{7.3} & 43.0\sd{8.3} & 34.0\sd{12.7} & 32.4\sd{6.6} & 25.6\sd{2.1} & 27.7\sd{4.2} & 17.1\sd{2.3} & 26.4\sd{4.6} \\
    \textsc{pca} & \textsc{last} & \textsc{conc} & \textsc{conc} & 62.6\sd{6.1} & 60.1\sd{3.5} & 56.2\sd{3.9} & 62.6\sd{2.5} & 50.9\sd{8.6} & 52.4\sd{2.0} & 41.3\sd{7.2} & 49.1\sd{5.1} & 13.7\sd{1.3} & 14.3\sd{0.6} & 18.7\sd{2.7} & 32.9\sd{1.9} \\
    \textsc{pca} & \textsc{last} & \textsc{func} & \textsc{neut} & 38.3\sd{5.9} & 35.5\sd{3.7} & 38.1\sd{3.7} & 38.8\sd{5.9} & 25.2\sd{1.7} & 20.3\sd{1.2} & 17.5\sd{1.0} & 19.4\sd{1.2} & 26.8\sd{1.4} & 20.7\sd{2.3} & 30.5\sd{8.5} & 30.2\sd{3.9} \\
    \textsc{pca} & \textsc{last} & \textsc{func} & \textsc{conc} & 42.7\sd{5.5} & 39.7\sd{7.8} & 36.1\sd{9.0} & 36.7\sd{5.5} & 30.5\sd{2.6} & 26.4\sd{9.6} & 30.3\sd{7.8} & 32.3\sd{7.1} & 16.3\sd{4.3} & 9.6\sd{1.5} & 13.8\sd{3.8} & 17.0\sd{2.3} \\
    \addlinespace
    \textsc{pca} & \textsc{avg} & \textsc{neut} & \textsc{neut} & 64.5\sd{5.4} & 65.7\sd{11.5} & 63.4\sd{7.8} & 61.8\sd{6.3} & 34.6\sd{12.7} & 45.8\sd{13.8} & 32.9\sd{16.2} & 49.5\sd{10.4} & 25.1\sd{3.5} & 21.2\sd{5.6} & 22.3\sd{3.9} & 32.2\sd{8.9} \\
    \textsc{pca} & \textsc{avg} & \textsc{neut} & \textsc{conc} & 49.2\sd{6.8} & 40.9\sd{6.0} & 42.4\sd{4.8} & 48.8\sd{8.0} & 49.5\sd{2.3} & 41.2\sd{2.5} & 32.1\sd{6.6} & 29.4\sd{5.5} & 23.0\sd{7.3} & 24.2\sd{10.9} & 20.2\sd{4.7} & 25.7\sd{3.0} \\
    \textsc{pca} & \textsc{avg} & \textsc{conc} & \textsc{neut} & 41.3\sd{7.5} & 52.0\sd{4.4} & 41.3\sd{5.9} & 56.9\sd{4.4} & 44.5\sd{8.4} & 49.3\sd{10.4} & 44.9\sd{16.3} & 41.0\sd{18.3} & 26.4\sd{1.2} & 27.0\sd{3.5} & 28.0\sd{11.5} & 29.3\sd{6.7} \\
    \textsc{pca} & \textsc{avg} & \textsc{conc} & \textsc{conc} & 61.2\sd{5.1} & 59.9\sd{3.9} & 58.7\sd{6.7} & 63.1\sd{1.8} & 57.0\sd{7.1} & \textbf{54.2}\sd{7.1} & 44.7\sd{5.4} & 39.7\sd{8.8} & 13.0\sd{0.9} & 12.9\sd{1.1} & 16.7\sd{6.0} & 23.6\sd{1.9} \\
    \textsc{pca} & \textsc{avg} & \textsc{func} & \textsc{neut} & 61.4\sd{5.3} & 66.0\sd{2.6} & 63.3\sd{5.9} & 65.2\sd{6.2} & 35.5\sd{14.0} & 38.3\sd{11.2} & 31.2\sd{11.6} & \textbf{53.8}\sd{12.7} & \textbf{33.3}\sd{3.7} & 24.2\sd{5.5} & 22.0\sd{2.9} & 29.5\sd{7.3} \\
    \textsc{pca} & \textsc{avg} & \textsc{func} & \textsc{conc} & 26.0\sd{5.3} & 41.8\sd{6.1} & 36.9\sd{6.5} & 48.2\sd{4.6} & 27.1\sd{10.1} & 38.9\sd{7.2} & 26.9\sd{8.6} & 27.1\sd{7.2} & 10.7\sd{2.2} & 13.7\sd{2.7} & 14.6\sd{4.1} & 17.2\sd{2.9} \\
    \addlinespace
    \textsc{md} & \textsc{last} & \textsc{neut} & \textsc{neut} & 48.5\sd{3.9} & 54.1\sd{10.0} & 48.6\sd{6.1} & 48.6\sd{7.2} & 23.1\sd{1.8} & 23.2\sd{1.4} & 20.4\sd{1.5} & 19.1\sd{1.2} & 31.5\sd{1.1} & 24.3\sd{3.1} & 25.5\sd{4.4} & 22.9\sd{2.2} \\
    \textsc{md} & \textsc{last} & \textsc{neut} & \textsc{conc} & 50.2\sd{3.6} & 58.4\sd{8.5} & 51.7\sd{8.2} & 49.6\sd{5.4} & 32.5\sd{2.8} & 19.9\sd{8.3} & 27.8\sd{6.9} & 28.6\sd{7.3} & 27.4\sd{7.5} & 11.0\sd{4.4} & 14.1\sd{3.9} & 28.9\sd{4.9} \\
    \textsc{md} & \textsc{last} & \textsc{conc} & \textsc{neut} & 37.3\sd{4.1} & 54.8\sd{6.1} & 48.0\sd{7.3} & 50.6\sd{3.8} & 22.0\sd{6.2} & 36.1\sd{10.7} & 38.0\sd{7.7} & 28.0\sd{8.2} & 27.5\sd{3.5} & \textbf{30.1}\sd{4.5} & 20.2\sd{6.2} & 26.1\sd{5.1} \\
    \textsc{md} & \textsc{last} & \textsc{conc} & \textsc{conc} & 66.0\sd{6.3} & 68.7\sd{4.8} & 63.0\sd{3.1} & 65.1\sd{2.0} & 50.7\sd{12.3} & 52.1\sd{2.0} & 43.9\sd{2.3} & 41.2\sd{8.3} & 14.9\sd{2.0} & 14.1\sd{0.9} & 18.4\sd{3.2} & 32.8\sd{2.0} \\
    \textsc{md} & \textsc{last} & \textsc{func} & \textsc{neut} & 46.1\sd{4.9} & 49.0\sd{7.5} & 47.6\sd{4.3} & 47.6\sd{7.6} & 25.6\sd{1.8} & 20.7\sd{1.2} & 17.7\sd{0.9} & 19.7\sd{1.4} & 27.8\sd{2.2} & 19.4\sd{1.7} & 24.6\sd{4.3} & 27.6\sd{4.3} \\
    \textsc{md} & \textsc{last} & \textsc{func} & \textsc{conc} & 39.7\sd{8.2} & 46.1\sd{10.2} & 48.4\sd{6.9} & 54.3\sd{2.6} & 31.6\sd{3.2} & 28.6\sd{9.9} & 29.6\sd{9.4} & 31.1\sd{7.4} & 14.3\sd{3.4} & 9.5\sd{1.2} & 9.5\sd{0.9} & 17.3\sd{2.1} \\
    \addlinespace
    \textsc{md} & \textsc{avg} & \textsc{neut} & \textsc{neut} & \textbf{73.5}\sd{4.9} & \textbf{77.4}\sd{2.6} & \textbf{71.2}\sd{4.2} & \textbf{74.8}\sd{5.1} & 36.3\sd{2.8} & 52.2\sd{7.0} & 41.3\sd{7.7} & 41.1\sd{4.3} & 24.6\sd{4.2} & 16.3\sd{3.9} & 18.7\sd{2.3} & 28.4\sd{3.8} \\
    \textsc{md} & \textsc{avg} & \textsc{neut} & \textsc{conc} & 50.4\sd{6.7} & 49.8\sd{7.8} & 57.9\sd{8.3} & 50.6\sd{8.3} & 29.8\sd{5.3} & 45.6\sd{4.2} & 30.2\sd{3.1} & 31.0\sd{2.1} & 20.6\sd{9.4} & 15.8\sd{1.7} & 20.9\sd{10.8} & 28.3\sd{3.6} \\
    \textsc{md} & \textsc{avg} & \textsc{conc} & \textsc{neut} & 50.7\sd{7.7} & 54.0\sd{4.2} & 62.1\sd{7.7} & 56.5\sd{6.6} & 44.5\sd{10.9} & 42.3\sd{7.0} & \textbf{57.9}\sd{6.9} & 47.0\sd{5.3} & 28.1\sd{3.8} & 29.2\sd{4.1} & \textbf{31.1}\sd{9.6} & 28.9\sd{7.1} \\
    \textsc{md} & \textsc{avg} & \textsc{conc} & \textsc{conc} & 64.6\sd{4.8} & 66.8\sd{4.5} & 64.7\sd{3.8} & 67.0\sd{1.9} & \textbf{60.0}\sd{4.8} & 51.0\sd{11.1} & 42.0\sd{3.9} & 34.2\sd{3.3} & 13.4\sd{1.2} & 13.1\sd{1.2} & 12.8\sd{3.6} & 22.7\sd{2.7} \\
    \textsc{md} & \textsc{avg} & \textsc{func} & \textsc{neut} & 68.9\sd{3.8} & 72.4\sd{4.1} & 69.6\sd{3.9} & 70.6\sd{4.5} & 29.0\sd{11.7} & 40.1\sd{4.0} & 35.6\sd{5.9} & 50.9\sd{13.8} & 30.8\sd{1.9} & 21.5\sd{6.1} & 21.7\sd{4.3} & 31.0\sd{5.8} \\
    \textsc{md} & \textsc{avg} & \textsc{func} & \textsc{conc} & 46.0\sd{4.4} & 46.7\sd{5.9} & 57.4\sd{7.0} & 55.9\sd{8.0} & 28.5\sd{4.6} & 39.0\sd{12.0} & 32.6\sd{6.3} & 34.6\sd{7.8} & 12.7\sd{2.3} & 13.8\sd{2.7} & 11.3\sd{1.1} & 16.0\sd{0.8} \\
    \bottomrule
  \end{tabular}%
  }
\end{table*}

\begin{table*}[t]
  \centering
  \caption{RQ1 configuration sweep on \he; columns and conventions as in \Cref{tab:rq1-bcb}. All numbers are percentages.}
  \label{tab:rq1-he}
  \scriptsize
  \setlength{\tabcolsep}{2pt}
  \resizebox{\textwidth}{!}{%
  \begin{tabular}{@{}llll *{12}{c}@{}}
    \toprule
     & & & & \multicolumn{4}{c}{in-distribution} & \multicolumn{4}{c}{OOD: \mbpp} & \multicolumn{4}{c}{OOD: synthetic} \\
    \cmidrule(lr){5-8}\cmidrule(lr){9-12}\cmidrule(lr){13-16}
    method & loc & fit & val/test & Mistral & Qwen & CodeLlama & OpenCoder & Mistral & Qwen & CodeLlama & OpenCoder & Mistral & Qwen & CodeLlama & OpenCoder \\
    \midrule
    \multicolumn{16}{@{}l}{\textit{Configuration of prior work~\cite{llm-correctness-icse2026}}}\\
    \textsc{pca} & \textsc{last} & \textsc{neut} & \textsc{conc} & 32.3\sd{3.8} & 56.0\sd{14.3} & 34.5\sd{9.1} & 42.2\sd{8.8} & 30.7\sd{3.2} & 27.3\sd{7.9} & 33.3\sd{10.3} & 22.9\sd{4.3} & 34.2\sd{6.0} & 53.4\sd{4.9} & 36.6\sd{11.0} & 32.3\sd{17.3} \\
    \midrule
    \multicolumn{16}{@{}l}{\textit{Full configuration sweep}}\\
    \textsc{pca} & \textsc{last} & \textsc{neut} & \textsc{neut} & 64.0\sd{4.0} & 68.9\sd{8.0} & 66.5\sd{4.7} & 65.5\sd{6.8} & 54.7\sd{4.5} & 78.1\sd{3.0} & \textbf{67.6}\sd{3.1} & 71.0\sd{2.7} & \textbf{64.4}\sd{5.8} & 57.3\sd{6.9} & 64.5\sd{4.1} & 69.7\sd{3.8} \\
    \textsc{pca} & \textsc{last} & \textsc{conc} & \textsc{neut} & 43.6\sd{10.8} & 70.8\sd{7.8} & 39.9\sd{12.6} & 59.5\sd{12.9} & 25.3\sd{7.9} & 29.7\sd{8.4} & 29.7\sd{11.2} & 37.8\sd{6.2} & 34.3\sd{7.9} & 31.9\sd{17.5} & 52.7\sd{6.7} & 60.8\sd{15.5} \\
    \textsc{pca} & \textsc{last} & \textsc{conc} & \textsc{conc} & 33.5\sd{5.2} & 56.4\sd{9.3} & 42.7\sd{9.3} & 52.3\sd{6.6} & 34.5\sd{10.4} & 31.9\sd{15.8} & 30.1\sd{7.6} & 32.5\sd{14.5} & 33.0\sd{3.2} & 62.4\sd{1.9} & 35.8\sd{4.2} & 57.3\sd{2.4} \\
    \textsc{pca} & \textsc{last} & \textsc{func} & \textsc{neut} & 63.5\sd{3.3} & 69.0\sd{6.1} & 66.0\sd{3.7} & 66.0\sd{6.2} & 59.3\sd{5.5} & 79.0\sd{3.6} & 63.3\sd{4.2} & 71.3\sd{2.4} & 54.2\sd{10.6} & 68.8\sd{5.1} & 56.0\sd{17.0} & 26.6\sd{24.7} \\
    \textsc{pca} & \textsc{last} & \textsc{func} & \textsc{conc} & 30.5\sd{3.0} & 55.2\sd{13.9} & 33.0\sd{7.0} & 40.4\sd{9.2} & 32.0\sd{2.4} & 27.9\sd{7.8} & 32.2\sd{7.2} & 33.7\sd{8.4} & 30.7\sd{4.0} & 61.8\sd{3.6} & 31.0\sd{5.5} & 40.0\sd{8.4} \\
    \addlinespace
    \textsc{pca} & \textsc{avg} & \textsc{neut} & \textsc{neut} & 35.6\sd{5.5} & 42.1\sd{12.7} & 36.4\sd{6.7} & 51.6\sd{9.3} & 38.3\sd{11.9} & 46.5\sd{8.0} & 41.8\sd{6.8} & 53.7\sd{6.8} & 32.6\sd{8.9} & 30.1\sd{11.3} & 33.1\sd{9.8} & 29.3\sd{15.1} \\
    \textsc{pca} & \textsc{avg} & \textsc{neut} & \textsc{conc} & 34.5\sd{8.3} & 49.3\sd{9.9} & 30.0\sd{4.7} & 50.2\sd{11.1} & 28.2\sd{4.7} & 41.5\sd{11.9} & 33.3\sd{4.2} & 33.0\sd{8.7} & 31.6\sd{10.0} & 50.2\sd{11.9} & 32.3\sd{8.2} & 36.4\sd{16.7} \\
    \textsc{pca} & \textsc{avg} & \textsc{conc} & \textsc{neut} & 32.4\sd{9.7} & 49.5\sd{8.7} & 40.7\sd{9.0} & 49.4\sd{7.5} & 26.8\sd{16.0} & 40.5\sd{4.2} & 40.8\sd{4.3} & 51.8\sd{8.5} & 34.1\sd{3.5} & 29.6\sd{4.6} & 28.4\sd{3.4} & 35.2\sd{5.8} \\
    \textsc{pca} & \textsc{avg} & \textsc{conc} & \textsc{conc} & 34.4\sd{3.4} & 58.8\sd{8.3} & 35.5\sd{10.0} & 42.7\sd{4.7} & 30.5\sd{7.6} & 29.8\sd{19.6} & 33.2\sd{11.2} & 34.6\sd{16.6} & 33.2\sd{3.2} & 63.6\sd{2.1} & 30.7\sd{3.4} & 47.4\sd{3.4} \\
    \textsc{pca} & \textsc{avg} & \textsc{func} & \textsc{neut} & 44.3\sd{8.9} & 59.3\sd{14.4} & 44.7\sd{11.0} & 62.8\sd{6.7} & 41.8\sd{10.1} & 64.9\sd{6.5} & 52.5\sd{11.8} & 56.1\sd{7.2} & 35.1\sd{2.8} & 40.7\sd{9.0} & 36.7\sd{3.3} & 27.9\sd{9.8} \\
    \textsc{pca} & \textsc{avg} & \textsc{func} & \textsc{conc} & 31.8\sd{5.4} & 60.3\sd{8.5} & 36.7\sd{5.7} & 47.9\sd{14.1} & 34.2\sd{6.8} & 58.8\sd{4.7} & 30.7\sd{8.8} & 34.9\sd{10.7} & 31.2\sd{2.7} & 66.9\sd{3.1} & 25.3\sd{7.7} & 42.0\sd{7.6} \\
    \addlinespace
    \textsc{md} & \textsc{last} & \textsc{neut} & \textsc{neut} & 64.6\sd{3.0} & 74.5\sd{9.2} & 68.0\sd{3.3} & 65.4\sd{6.1} & 55.1\sd{3.6} & 79.1\sd{2.9} & 67.2\sd{3.2} & 71.0\sd{2.7} & 60.9\sd{8.3} & 59.5\sd{9.3} & \textbf{65.1}\sd{3.6} & \textbf{70.5}\sd{4.2} \\
    \textsc{md} & \textsc{last} & \textsc{neut} & \textsc{conc} & 34.1\sd{6.6} & 67.9\sd{5.8} & 36.4\sd{5.6} & 41.2\sd{10.2} & 31.2\sd{4.1} & 34.0\sd{9.6} & 31.9\sd{9.3} & 23.7\sd{4.5} & 35.7\sd{4.6} & 54.4\sd{5.8} & 28.5\sd{7.1} & 49.0\sd{3.9} \\
    \textsc{md} & \textsc{last} & \textsc{conc} & \textsc{neut} & 45.5\sd{12.7} & \textbf{79.4}\sd{10.5} & 50.7\sd{12.5} & 58.5\sd{12.0} & 29.8\sd{9.1} & 47.4\sd{10.8} & 45.1\sd{12.5} & 45.5\sd{14.2} & 22.6\sd{8.9} & 23.0\sd{5.5} & 44.4\sd{19.3} & 60.6\sd{14.7} \\
    \textsc{md} & \textsc{last} & \textsc{conc} & \textsc{conc} & 39.4\sd{6.3} & 61.3\sd{8.2} & 49.4\sd{6.0} & 54.0\sd{4.5} & 44.9\sd{4.2} & 51.4\sd{3.8} & 40.8\sd{9.6} & 37.0\sd{16.1} & 33.1\sd{3.0} & 62.5\sd{2.0} & 35.5\sd{4.7} & 57.5\sd{2.5} \\
    \textsc{md} & \textsc{last} & \textsc{func} & \textsc{neut} & \textbf{64.7}\sd{4.5} & 68.7\sd{8.3} & \textbf{69.8}\sd{1.6} & 65.0\sd{5.2} & \textbf{59.9}\sd{5.0} & \textbf{79.1}\sd{3.2} & 63.0\sd{4.0} & \textbf{71.3}\sd{2.4} & 53.6\sd{5.6} & \textbf{71.7}\sd{3.1} & 63.8\sd{3.1} & 36.3\sd{25.7} \\
    \textsc{md} & \textsc{last} & \textsc{func} & \textsc{conc} & 32.4\sd{7.1} & 59.8\sd{10.6} & 35.4\sd{6.0} & 41.8\sd{7.6} & 32.6\sd{3.6} & 30.7\sd{6.7} & 38.9\sd{4.9} & 32.1\sd{8.5} & 31.3\sd{3.3} & 62.9\sd{2.7} & 24.5\sd{3.4} & 40.0\sd{5.9} \\
    \addlinespace
    \textsc{md} & \textsc{avg} & \textsc{neut} & \textsc{neut} & 49.8\sd{5.5} & 58.2\sd{9.7} & 51.6\sd{7.3} & 63.7\sd{8.8} & 32.3\sd{7.8} & 50.4\sd{4.3} & 49.5\sd{10.2} & 47.0\sd{3.3} & 35.1\sd{5.8} & 33.4\sd{11.4} & 33.3\sd{4.2} & 29.0\sd{13.9} \\
    \textsc{md} & \textsc{avg} & \textsc{neut} & \textsc{conc} & 29.9\sd{3.3} & 68.4\sd{13.4} & 32.1\sd{8.0} & 49.3\sd{10.0} & 26.9\sd{6.5} & 32.6\sd{12.6} & 37.9\sd{8.0} & 33.2\sd{5.8} & 34.8\sd{4.4} & 57.2\sd{6.4} & 32.6\sd{6.5} & 25.3\sd{14.5} \\
    \textsc{md} & \textsc{avg} & \textsc{conc} & \textsc{neut} & 34.5\sd{7.9} & 56.8\sd{5.1} & 40.9\sd{7.4} & 49.3\sd{10.8} & 36.4\sd{11.6} & 42.6\sd{3.5} & 41.9\sd{8.4} & 46.5\sd{15.5} & 33.2\sd{5.5} & 27.0\sd{4.0} & 25.3\sd{5.3} & 33.8\sd{6.3} \\
    \textsc{md} & \textsc{avg} & \textsc{conc} & \textsc{conc} & 45.0\sd{9.0} & 63.6\sd{11.4} & 47.8\sd{3.8} & 54.3\sd{8.5} & 43.6\sd{8.6} & 28.4\sd{12.0} & 47.5\sd{4.5} & 26.5\sd{4.7} & 32.7\sd{3.7} & 63.7\sd{2.3} & 30.9\sd{3.9} & 47.1\sd{3.3} \\
    \textsc{md} & \textsc{avg} & \textsc{func} & \textsc{neut} & 48.8\sd{5.7} & 64.0\sd{12.5} & 49.8\sd{6.0} & \textbf{66.8}\sd{8.1} & 38.2\sd{9.0} & 71.5\sd{4.8} & 48.2\sd{10.6} & 58.5\sd{9.7} & 38.0\sd{5.4} & 45.3\sd{14.8} & 35.9\sd{4.0} & 36.4\sd{17.1} \\
    \textsc{md} & \textsc{avg} & \textsc{func} & \textsc{conc} & 30.7\sd{5.8} & 61.2\sd{8.6} & 35.0\sd{8.6} & 54.0\sd{10.1} & 29.8\sd{5.0} & 46.0\sd{11.6} & 35.7\sd{6.6} & 37.7\sd{11.2} & 32.1\sd{3.1} & 66.9\sd{3.0} & 24.1\sd{4.7} & 45.0\sd{3.8} \\
    \bottomrule
  \end{tabular}%
  }
\end{table*}

\begin{table}[t]
  \centering
  \caption{ANOVA for RQ1. $\eta^2$ = variance share in \val\ from a factor
  (main effect) or an interaction. Panels~(1)--(4) follow
  Sections \ref{sec:rq1-general}--\ref{sec:rq1-construction-data-model}}
  \label{tab:rq1-nav}
  \footnotesize
  \setlength{\tabcolsep}{4pt}
  \begin{tabular}{l rl rl}
    \toprule
    \multicolumn{3}{@{}l}{\textit{(1) General View}}        & \multicolumn{2}{r}{$\eta^2$}\\
    \multicolumn{5}{@{}l}{\quad\textit{main effects}}\\
    \multicolumn{3}{@{}l}{\quad\phantom{0}2.~model identity}              & \multicolumn{2}{r}{14\%}\\
    \multicolumn{3}{@{}l}{\quad\phantom{0}6.~testing framing (test)}      & \multicolumn{2}{r}{\phantom{0}7\%}\\
    \multicolumn{3}{@{}l}{\quad\phantom{0}7.~construction method}         & \multicolumn{2}{r}{\phantom{0}7\%}\\
    \multicolumn{3}{@{}l}{\quad22.~benchmark}                             & \multicolumn{2}{r}{${<}1\%$}\\
    \multicolumn{3}{@{}l}{\quad24.~read-out location (loc)}               & \multicolumn{2}{r}{${<}1\%$}\\
    \multicolumn{3}{@{}l}{\quad28.~fitting framing (fit)}                 & \multicolumn{2}{r}{${<}1\%$}\\
    \multicolumn{5}{@{}l}{\quad\textit{interactions}}\\
    \multicolumn{3}{@{}l}{\quad\phantom{0}1.~fit\,$\times$\,test framing}                      & \multicolumn{2}{r}{15\%}\\
    \multicolumn{3}{@{}l}{\quad\phantom{0}3.~loc\,$\times$\,benchmark}                         & \multicolumn{2}{r}{12\%}\\
    \multicolumn{3}{@{}l}{\quad\phantom{0}4.~loc\,$\times$\,test framing\,$\times$\,benchmark} & \multicolumn{2}{r}{11\%}\\
    \multicolumn{3}{@{}l}{\quad\phantom{0}5.~benchmark\,$\times$\,model}                       & \multicolumn{2}{r}{\phantom{0}7\%}\\
    \addlinespace
    \multicolumn{3}{@{}l}{\quad any term involving benchmark} & \multicolumn{2}{r}{47\%}\\
    \midrule
    & \multicolumn{2}{c}{\bcb} & \multicolumn{2}{c}{\he} \\
    \cmidrule(lr){2-3}\cmidrule(lr){4-5}
    & $\eta^2$ & best & $\eta^2$ & best \\
    \midrule
    \multicolumn{5}{@{}l}{\textit{(2) Within each benchmark}}\\
    read-out location    & 20\%     & avg         & \phantom{0}6\%  & last        \\
    construction method  & 14\%     & \textsc{md} & \phantom{0}3\%  & \textsc{md} \\
    fitting framing      & \phantom{0}5\% & conc  & \phantom{0}1\%  & func        \\
    testing framing      & ${<}1\%$ & ---         & 20\%            & neut        \\
    \addlinespace
    model identity       & 11\%     & agree       & 51\%            & Qwen        \\
    \addlinespace
    \quad fit\,$\times$\,test framing & 26\% & & \phantom{0}7\% & \\
    \quad loc\,$\times$\,test framing & 15\% & & \phantom{0}8\% & \\
    \midrule
    \multicolumn{3}{@{}l}{\textit{(3) Out-of-distribution fitting}} & \multicolumn{2}{r}{\textit{mean acc.}}\\
    \quad in-distribution & \multicolumn{2}{c}{52\%} & \multicolumn{2}{c}{50\%} \\
    \quad \mbpp           & \multicolumn{2}{c}{35\%} & \multicolumn{2}{c}{43\%} \\
    \quad synthetic       & \multicolumn{2}{c}{22\%} & \multicolumn{2}{c}{42\%} \\
    \midrule
    \multicolumn{5}{@{}l}{\textit{(4) Construction, data, and model}}\\
    \quad construction & \multicolumn{2}{c}{16\%} & \multicolumn{2}{c}{42\%} \\
    \quad fitting data & \multicolumn{2}{c}{75\%} & \multicolumn{2}{c}{18\%} \\
    \quad model        & \multicolumn{2}{c}{\phantom{0}9\%} & \multicolumn{2}{c}{40\%} \\
    \bottomrule
  \end{tabular}
\end{table}

\subsubsection{General View}\label{sec:rq1-general}
We take the broadest view first: we pool both benchmarks (\bcb\ and \he) and rank
the choices by how much of the variance in \val\ each one carries. This view is
restricted to in-distribution fitting; the fitting source is examined on its own
later (\Cref{tab:rq1-nav}, panel~3). The restriction is needed because
\emph{in-distribution} does not name a single fitting source: it is \bcb\ data when
we test on \bcb\ and \he\ data when we test on \he, whereas the out-of-distribution
sources --- \mbpp\ and the synthetic set --- are the same whichever benchmark we
test on. That is, the benchmark and the in-distribution source change together and
cannot be told apart.

We find that the resulting decomposition (\Cref{tab:rq1-nav}, panel~1) spreads the variance across many terms
rather than concentrating it in one. The two largest shares, $15\%$ and $14\%$, are
\begin{enumerate*}[label=(\arabic*)]
  \item the combination of fitting and testing framing, and
  \item the model.
\end{enumerate*}
The benchmark on its own is negligible --- under $1\%$ --- but it frequently appears together with other interaction terms: with the
read-out location ($12\%$), with the location and testing framing together ($11\%$),
and with the model ($7\%$). Summed together, the terms involving the benchmark account for about
$47\%$ of the variance (final row in \Cref{tab:rq1-nav}, panel~1). So the benchmark alone
does not set the accuracy level; it sets \emph{which other design choices matter}. The
read-out location is the clearest example: on its own it explains under $1\%$, but
paired with the benchmark it explains $12\%$, because the location that performs best on
one benchmark is worse on the other (in panel~2 its \emph{best}
level flips between the response average and the last token). This pattern sets up the next step of the analysis. If the benchmark decides which level of a choice works best, then a ranking computed over the pooled benchmarks averages away exactly that information: a choice can carry little variance in the pool not because it is unimportant, but because its opposite effects on the two benchmarks cancel, as they do for the read-out location. Therefore, the pooled view can say that the benchmark shapes the other choices, but not which setting to prefer on a given benchmark; for that, we next examine each benchmark on its own (\Cref{tab:rq1-nav}, panel~2).

\subsubsection{Within each benchmark}
Examined on its own, each benchmark ranks the choices differently (\Cref{tab:rq1-nav}, panel~2).
On \bcb\ the largest share is not a single choice but the fitting and testing framing
acting together ($26\%$), followed by the read-out location ($20\%$) and that same
location combined with the testing framing ($15\%$). The location matters, but it
cannot be set on its own: the framing it is read under changes its effect. Concept
fitting illustrates this: on \bcb, a \meandiff\ direction fit under the concept framing
and read at the last token scores $63$--$69\%$ across the four models when the testing
framing is also concept, but $37$--$55\%$ under a neutral one (\Cref{tab:rq1-bcb}).

Construction method is the exception, the one choice that behaves the same way whatever
the rest is set to. It is a main effect with no large interaction beside it ($14\%$ on
\bcb, $3\%$ on \he), and its better level is \meandiff\ on both benchmarks; it is the
only construction choice whose preferred setting never reverses.

Every other choice depends on the benchmark, and the read-out location is the clearest
case. On \bcb\ the response average wins: a neutral \meandiff\ direction reaches
$71$--$77\%$ across the four models when averaged over the response, against
$48$--$54\%$ at the last token (\Cref{tab:rq1-bcb}). On \he\ the last token wins,
$65$--$75\%$ against $50$--$64\%$ for the same direction (\Cref{tab:rq1-he}). These two
opposite preferences are what cancelled in the pooled view --- averaging over the
benchmarks left the location under $1\%$ in panel~1 --- and they are why its
\emph{best} level flips between the two benchmarks in panel~2.

The models follow the same benchmark-dependent pattern. On \bcb\ they agree: model identity accounts for only
$11\%$ of the variance, and the best configuration lands within a few points across the
four (panel~2). On \he\ they disagree sharply: model identity accounts for
$51\%$, the largest single share on either benchmark, and the best per-model
configuration ranges from $42\%$ (Mistral) to $62\%$ (Qwen) (\Cref{tab:rq1-he}). A
further $20\%$ on \he\ is the testing framing on its own, which favours the neutral
prompt (panel~2). On \he, a configuration that is best for one model need
not be best for another.

\subsubsection{Out-of-distribution fitting}
We now return to the factor held out of the general view (\Cref{sec:rq1-general}): the source of the fitting
data. So far, every direction was fit on its own benchmark's fitting data. Now, we fit
it on data from \mbpp\ or the synthetic set, while still
testing on the same benchmark (the right two blocks of \Cref{tab:rq1-bcb} and
\Cref{tab:rq1-he}). Panel~3 of \Cref{tab:rq1-nav} reports the mean \val\ for each source.
In-distribution fitting is best on both benchmarks, but both decrease when fitting out-of-distribution. On \he\ it is a small decrease: mean accuracy goes from $50\%$ to
$43\%$ on \mbpp\ and $42\%$ on the synthetic set, a decrease of 7 and 8 points, respectively. On \bcb\
the decrease is larger: from $52\%$ to $35\%$ on \mbpp\ and to $22\%$ on the synthetic set, a
decrease of 17 and 30 points, respectively. The fitting source therefore matters on both benchmarks,
but considerably more on \bcb\ than on \he. At $22\%$, the synthetic-fit mean on \bcb\ has dropped to
the $25\%$ chance level of the 1-of-4 selection: a poorly matched fitting source does not merely weaken
the correctness signal but can reduce it to chance.

\subsubsection{Construction, data, and model}\label{sec:rq1-construction-data-model}
The decomposition so far has more terms than can be compared across the two benchmarks
at a glance. Producing a direction takes three different
components, and every factor in our work can be attributed to one of them:
\begin{enumerate*}[label=(\arabic*)]
  \item how the direction is \emph{constructed}: the method, read-out location, and
    framing;
  \item what \emph{data} it is fit on: in-distribution and out-of-distribution;
  \item which \emph{model} it is read from: 4 LLMs which are neither a construction choice nor part of the data.
\end{enumerate*}
These three groups comprise everything, and panel~4 of \Cref{tab:rq1-nav} reports them per benchmark as before. On \bcb\ the fitting data carries $75\%$ of the variance, construction $16\%$, and the model $9\%$. On \he\ the order is different: construction
carries $42\%$, the model $40\%$, and the fitting data $18\%$. The kind of choice that
carries the most variance is therefore not the same on the two benchmarks: on \bcb\ it
is the data the direction is fit on, and on \he\ it is how the direction is built and
which model it is read from.

\begin{takeawaybox}
There is no single best configuration for capturing a code correctness direction.
Only the construction method generalises: \meandiff\ beats PCA. Read-out location,
framing, and model are not consistent across benchmarks, so which \emph{kind} of
choice matters most is benchmark-dependent: on \bcb\ the fitting-data source accounts
for most of the variance ($75\%$), while on \he\ it splits between construction
($42\%$) and model ($40\%$).
\end{takeawaybox}

\section{RQ2: Does controlling for differences between correct/incorrect programs improve the quality of the correctness direction?}
\label{sec:rq2}

\Cref{tab:rq2-transfer} reports the accuracy of correctness directions fit on
\emph{controlled pairs}: pairs whose incorrect member is constructed from the
correct one rather than drawn from a plausible failing attempt (\Cref{subsec:priorwork}). It covers both variants on both benchmarks --- M-Only,
whose incorrect member is a single-fault mutation of the correct solution, and
MC, which adds a behaviour-preserving rewrite on top of that fault
(\Cref{sec:rq2-data}). Its upper block is an aggregated view of RQ1's results on the same test sets and differs from the lower blocks only in the fitting source, not the test set. Each lower block fits on one benchmark's controlled pairs and its last column tests on a held-out split of that same source. Each cell is \val{} with the best-layer accuracy in parentheses (\Cref{sec:setup-protocol}). By default, the layer is selected on a held-out split of the fitting pairs, while \Cref{sec:rq2-layer} also reports selecting it on a held-out split of the test benchmark, as RQ1 does in-distribution.

Recall that RQ1 found \meandiff{} to generally give the best directions across both
benchmarks, with no single best read-out location, framing, or model (\Cref{sec:rq1}).
Therefore, every direction in \Cref{tab:rq2-transfer} is built with \meandiff{} and averaged over the other choices.
As a check on how the direction is built, \Cref{tab:rq2-robust}
repeats the two same-benchmark fits with PCA and with paired sampling. We report M-Only in detail, then the MC variant to test
how adding controlled rewrites affects the results (\Cref{sec:rq2-confounds}).

\begin{table}[t]
  \centering
  \caption{RQ2 transfer: \meandiff\ direction fit on a benchmark's controlled pairs
  (M-Only/MC), test on \bcb, \he, and held-out split of those pairs. Cells are
  \val\ (best layer in parentheses), averaged over read-out locations, framings, and
  models. Top block is RQ1's \meandiff\ on the same test sets, differs only in
  fitting source. Dashes mark conditions not run; \textsuperscript{\dag}~fits on one
  benchmark, tests on the other.}
  \label{tab:rq2-transfer}
  \footnotesize
  \setlength{\tabcolsep}{3pt}
  \begin{tabular}{l ccc}
    \toprule
     & \multicolumn{3}{c}{tested on} \\
    \cmidrule(lr){2-4}
    fit source & \bcb{} & \he{} & held-out \\
    \midrule
    \multicolumn{4}{@{}l}{\textit{RQ1 sources}}\\
    \quad in-distribution & 56.7\,(61.6) & 52.8\,(61.7) & --- \\
    \quad \mbpp{}         & 35.2\,(49.2) & 44.4\,(57.7) & --- \\
    \quad synthetic       & 21.5\,(40.5) & 42.6\,(53.3) & --- \\
    \addlinespace
    \multicolumn{4}{@{}l}{\textit{fit on \bcb{} controlled pairs}}\\
    \quad M-Only & 19.6\,(36.8) & 29.8\,(43.1)\textsuperscript{\dag} & 83.3\,(86.4) \\
    \quad MC     & 19.0\,(36.7) & 30.0\,(43.0)\textsuperscript{\dag} & 81.7\,(84.3) \\
    \addlinespace
    \multicolumn{4}{@{}l}{\textit{fit on \he{} controlled pairs}}\\
    \quad M-Only & 22.2\,(38.9)\textsuperscript{\dag} & 47.6\,(57.3) & 71.6\,(78.9) \\
    \quad MC     & 21.3\,(37.1)\textsuperscript{\dag} & 48.5\,(57.9) & 71.1\,(78.5) \\
    \bottomrule
  \end{tabular}
\end{table}

\begin{table}[t]
  \centering
  \caption{RQ2 robustness (M-Only). \val{} for a direction fit on a benchmark's
  controlled pairs, tested in isolation (a held-out split of those pairs) and in
  transfer (the benchmark), across construction method and sampling. Isolation stays
  high and transfer low throughout, so neither accounts for the RQ1 gap; paired
  sampling separates held-out pairs at least as well yet transfers no better. MC
  behaves the same (\Cref{tab:rq2-transfer}).}
  \label{tab:rq2-robust}
  \footnotesize
  \setlength{\tabcolsep}{4pt}
  \begin{tabular}{ll cc cc}
    \toprule
     & & \multicolumn{2}{c}{\bcb} & \multicolumn{2}{c}{\he} \\
    \cmidrule(lr){3-4}\cmidrule(lr){5-6}
    method & sampling & isol. & transf. & isol. & transf. \\
    \midrule
    \meandiff{} & weighted & 83.3 & 19.6 & 71.6 & 47.6 \\
    \meandiff{} & paired   & 85.9 & 19.1 & 78.4 & 48.0 \\
    PCA         & weighted & 81.4 & 20.0 & 65.1 & 46.1 \\
    PCA         & paired   & 83.2 & 19.7 & 73.0 & 44.7 \\
    \bottomrule
  \end{tabular}
\end{table}

\subsubsection{M-Only pairs are separable in isolation}\label{sec:rq2-heldout}
Fit, validated, and tested on a held-out split of the same controlled pairs, the
M-Only direction reaches $83.3\%$ on \bcb\ and $71.6\%$ on \he\ ($86.4\%$ and
$78.9\%$ at the best layer). Separation this high on unseen pairs means the pairs carry a
consistent signal that is learnable: had the mutations produced malformed or degenerate
programs, a direction fit on one split of the pairs could not separate another. For
context, these values even sit above each benchmark's own in-distribution fit
($56.7\%$ and $52.8\%$), although the two settings answer different questions ---
one separates a solution from its mutants, the other ranks the benchmark's
candidate implementations. Still, the held-out accuracy is lower on
\he\ than on \bcb{}. This may reflect the smaller pool of \he\ mutants ($4{,}566$ against $27{,}928$ from \bcb{}). Whether this in-isolation signal is the
correctness direction, and whether it carries over in transfer to the benchmarks, is the core question for the rest of this section.

\subsubsection{Testing on \bcb}
We now change only the test set: the direction is again fit and validated on
\bcb{}'s M-Only pairs, but instead of scoring held-out pairs of the same
construction, it faces \bcb{}'s own 1-of-4 selection task over candidate
implementations, the same test RQ1's directions face. There, the
accuracy reaches $19.6\%$, against $56.7\%$ for RQ1's in-distribution fit on the
same \bcb\ tasks --- a decrease of $37.1$ points. The M-Only fit also sits below
every RQ1 source, including \mbpp\ ($35.2\%$) and the synthetic set ($21.5\%$),
neither of which shares the \bcb\ tasks that the controlled pairs keep. Since these
pairs are highly separable in isolation (\Cref{sec:rq2-heldout}), the gap to the
in-distribution fit is a transfer failure: the signal that tells a solution from its
mutation does not carry over to ranking \bcb{}'s plausible failing attempts.

\subsubsection{Testing on \he}
Fit and validated on \he{}'s M-Only pairs and tested on \he{}, the
accuracy reaches $47.6\%$, against $52.8\%$ for RQ1's in-distribution fit on \he\
--- a decrease of $5.2$ points, far smaller than the $37.1$-point decrease on
\bcb{}. As such, on \he{}, isolating the bug-causing change leaves the direction closer to the in-distribution fit, while on \bcb\ the equivalent isolation lowers accuracy by more than half. This split by benchmark follows the pattern of RQ1, where the effect of each construction choice also depended on the benchmark (\Cref{sec:rq1}).

\subsubsection{Crossing benchmarks}
Fitting on one benchmark's controlled pairs and testing on the other (the daggered
cells) gives low accuracy in both directions, but these change the benchmark
between fitting and testing and so do not isolate the fault's effect. The two
crossings are not symmetric: fitting on \bcb\ and testing on \he\ ($29.8\%$) does
somewhat better than the reverse ($22.2\%$). Were the gap purely a format mismatch
between benchmarks, we might expect the two crossings to be closer. The asymmetry may instead reflect differences in task complexity: the higher difficulty of \bcb\ enables a higher-quality direction that then transfers better to \he\ than the reverse.

\subsubsection{Direction quality}
On \bcb{}, the M-Only fit's validation-selected and best-layer accuracies differ
widely ($19.6\%$ and $36.8\%$; a $17.2$-point gap), far more than for the
in-distribution fit ($56.7\%$ and $61.6\%$; a $4.9$-point gap). The decrease is not
merely a matter of poor layer selection: even at the best layer, the
M-Only direction ($36.8\%$) is $24.8$ points below the in-distribution fit
($61.6\%$), so the direction itself is worse, not merely read at a suboptimal
layer. On \he{}, no such gap appears: at the best layer the M-Only direction
($57.3\%$) is only $4.4$ points below the in-distribution fit ($61.7\%$). The weaker
direction is therefore specific to \bcb{}.

\subsubsection{Selecting the layer on the benchmark}\label{sec:rq2-layer}
The wide $17.2$-point gap between the M-Only fit's validation-selected and
best-layer accuracies on \bcb\ reflects a second cost, this time from layer
selection itself: the layer that best separates the controlled pairs need not
be the layer that best ranks the benchmark's candidates. We quantify this cost by
selecting the layer on a held-out split of the test
benchmark --- as RQ1's in-distribution fits do --- rather than on the controlled
pairs. Doing so raises the \bcb\ M-Only fit from $19.6\%$ to $32.9\%$, most of the
way to its $36.8\%$ oracle: about $13$ of the
$17.2$-point gap is a layer-selection mismatch between the controlled pairs and
\bcb{}, the rest a genuinely weaker direction. On \he{}, the same swap barely moves
the M-Only fit ($47.6\%$ to $48.6\%$): validating on \he{}'s controlled pairs already
selects essentially the same layer as validating on the benchmark itself, so there is
no mismatch to recover, and the added cost is again specific to \bcb{}.

\subsubsection{Adding controlled confounds}\label{sec:rq2-confounds}
The MC variant applies behaviour-preserving rewrites, e.g. variable renaming, so that the two implementations differ in surface form while the fault stays fixed (\Cref{sec:rq2-data}). Across every measure, MC
tracks M-Only. It is comparably separable in isolation ($81.7\%$ on \bcb\ and
$71.1\%$ on \he{}, against $83.3\%$ and $71.6\%$), its transfer accuracy is just as
low ($19.0\%$ on \bcb\ and $48.5\%$ on \he{}, against $19.6\%$ and $47.6\%$), and it
benefits the same way from selecting the layer on the benchmark ($19.0\%$ to
$32.7\%$ on \bcb{}). Therefore, adding controlled surface variation neither closes the
\bcb\ gap to the in-distribution fit nor disturbs the \he\ result: the captured
direction does not seem to depend on these surface features.

\subsubsection{Robustness to construction and sampling}
The gap to RQ1's in-distribution fit is not an artefact of how the direction is
built. \Cref{tab:rq2-robust} repeats each same-benchmark fit with PCA in place of
\meandiff{} and with paired in place of weighted sampling, reporting both isolation
and transfer. In transfer the accuracy stays within a narrow band on each benchmark
--- roughly $19$--$20\%$ on \bcb\ and $45$--$49\%$ on \he\ across all four
method--sampling combinations (M-Only shown; MC is comparable, \Cref{tab:rq2-transfer}). In isolation every combination
stays highly separable, and paired sampling, which fits on every available pair
rather than one per task, separates the held-out pairs better than weighted
sampling (M-Only $85.9\%$ against $83.3\%$ on \bcb{}, $78.4\%$ against $71.6\%$ on
\he{}). That the larger, better-separated paired fitting set still transfers no
better reinforces that the in-isolation signal does not lead to a more capable
correctness direction (\Cref{sec:rq2-heldout}).

\begin{takeawaybox}[frametitle={RQ2: Key takeaway}]
Fitting on program pairs that differ only by the bug-causing change yields a
direction that separates such pairs well, yet ranks the benchmarks'
candidate implementations worse than RQ1's in-distribution fit --- slightly
worse on \he, substantially worse on \bcb. On \bcb{}, the M-Only fit reaches
$19.6\%$, against $56.7\%$ on the same tasks; on \he\
it reaches $47.6\%$, against $52.8\%$, a much smaller decrease. MC barely changes results ($19.0\%$ and
$48.5\%$), and construction method and sampling do not as well (\Cref{tab:rq2-robust}). In isolation M-Only is separable ($83.3\%$ and $71.6\%$, MC comparable), so the decrease points to a
property of the captured direction rather than malformed pairs.
\end{takeawaybox}

\section{Discussion}
\label{sec:discussion}

\subsection{The nature of the captured direction}
The captured code correctness direction is not a single property of a
model: its accuracy, and which configuration recovers it best, depend on how the
direction is constructed and what it is contrasted against. RQ1 varies the
construction: one choice generalises across both benchmarks --- \meandiff\ beats
PCA --- but the rest do not, and which \emph{kind} of choice matters most is itself
decided by the benchmark (\Cref{sec:rq1}). RQ2 varies the contrast: a direction fit
on a benchmark's controlled pairs separates a held-out split of them well ($83.3\%$
on \bcb\ and $71.6\%$ on \he, higher still under paired sampling), yet on \bcb\
that separability does not carry over to ranking the benchmark's own plausible
failing attempts ($19.6\%$, against $56.7\%$ for the in-distribution fit). Telling a
program from its own single-fault mutation and judging whether independent
candidates are correct are thus not the same property. Paired sampling makes this clearer:
fitting on every pair rather than one per task raises separability but
leaves transfer unchanged. This qualifies the broad reading of internal ``truth''
or ``correctness'' directions~\cite{azaria2023internal, marks2023geometry,
zou2023representation}: their identity and strength are properties of the
configuration and contrast set, not of the model alone.

\subsection{Controlling the fitting data}
We expected that removing the
incidental differences between a correct program and a plausible failing attempt
would sharpen the direction, leaving only the fault. Our results do
not bear this out. Isolating the bug-causing change does not improve transfer on
either benchmark, and on \bcb\ it lowers it substantially: the M-Only fit reaches
$19.6\%$ against $56.7\%$ for the in-distribution fit on the same tasks, below even
the out-of-distribution sources that share none of those tasks (\mbpp\ $35.2\%$,
synthetic $21.5\%$). On \he\ the same isolation costs only $5.2$ points
($47.6\%$ against $52.8\%$); as in RQ1, the benchmark sets the size of the effect.
Part of the \bcb\ gap is a layer-selection mismatch --- selecting the layer on the
benchmark recovers about $13$ of the $17.2$-point gap --- but even at
the oracle layer the mutation direction stays $24.8$ points below it, so a genuinely
weaker direction remains.

We read the \bcb\ result as a mismatch between the contrast the direction is fit on
and the programs it later scores. Both controlled variants build the incorrect
member by perturbing the correct solution --- M-Only seeds a single fault, MC adds
behaviour-preserving rewrites --- whereas a benchmark's plausible failing attempts
are independent implementations, free to differ in algorithm and structure, not
only at the fault. That MC leaves \bcb\ transfer unchanged ($19.0\%$) indicates
the missing ingredient is not surface variety but this difference in kind: a
direction tuned to separate a solution from a perturbed copy of itself is not
thereby tuned to judge an independent attempt. The high separability
($83.3\%$) rules out malformed mutations; whether the injected faults resemble the
mistakes models make in practice is taken up in \Cref{sec:threats}.

\subsection{Practical use of the technique}
The code correctness signal is usable, but should be tuned per setting rather than
applied as a fixed recipe. First, construct the direction with \meandiff{}: it is the one
choice whose better setting never reverses, generally beating PCA on
both benchmarks (\Cref{sec:rq1}). Second, do not fix the read-out location or framing
in advance; the best location reverses between \he\ and \bcb, so select it per model
and benchmark on a validation split. The gap between validation-selected and
best-layer accuracy is itself a cost to report, not hide: on \bcb\ it is wide
for the mutation fit, and quoting only the oracle layer overstates what the probe
achieves. More generally, a single headline accuracy --- including prior work's
$41$--$63\%$ range~\cite{llm-correctness-icse2026} --- reflects one configuration
out of many, and our sweep shows that accuracy varies widely across them. That is, reporting the
validation-selected accuracy of the whole sweep describes what the technique achieves
without tying the claim to one configuration. The same variation explains the value of
approaches that do not fix the read-out at all: AutoProbe~\cite{vu2026autoprobe} learns
which hidden states to read per model instead of committing to a pre-selected layer and
token position. Our results quantify the accuracy decrease when that choice
is made poorly.

\section{Threats to Validity}
\label{sec:threats}

\textbf{Construct Validity.} Using unit tests as a proxy for
correctness may be incomplete and does not capture non-functional
aspects, \eg performance. Moreover, high probing accuracy alone is weak evidence
that the captured direction encodes correctness and not a correlated
feature~\cite{belinkov2022probing}: a sufficiently expressive probe can score
well by memorising the labels rather than by reading them from the hidden
states~\cite{hewitt2019control}; probes can score highly on a property the model does not
use~\cite{ravichander2021probing}, and prove unreliable for concept removal and
detection~\cite{elazar2021amnesic, kumar2022probing}. To mitigate this, our probes
are single linear directions, leaving them little capacity to memorise, and we do
not rely on a single configuration: we vary the construction method, framing, and
read-out location, select the layer on a validation split, and test on two
benchmarks with cross-validation. RQ2 further controls the fitting data with
mutations that isolate the fault and behaviour-preserving rewrites that vary only
surface form, checking whether the direction tracks correctness and not
incidental differences between programs.

\textbf{Internal Validity.} RQ2 showed that a direction fit on mutations did not
transfer to the benchmark's plausible failing attempts, which could reflect the
mutation process and not fault isolation alone. Our mutation pipeline only checks
types, not whether an edit resembles a mistake a developer or a model would
plausibly make, so some mutations may be unrealistic. Nonetheless, we use mutations
because isolating the bug-causing change requires pairs that differ only in a
single, localised fault, for which mutation is the standard way to
construct~\cite{just2014mutants}. More broadly, injected
faults need not resemble real ones and detectors trained on them transfer
poorly~\cite{patra2021semantic, richter2022learning, he2022distribution}, and
models complete code containing them less reliably~\cite{dinh2023large}. The
held-out mutant positive control reduces this concern:
the same direction separates held-out mutations, so the transfer result is
not an artefact of malformed or unlearnable mutations. Moreover, the confounds are
generated by \texttt{Qwen2.5-Coder-32B-Instruct}, which shares a model family with
one probed model (\texttt{Qwen2.5-Coder-7B-Instruct}), so rewrites in a familiar style could in principle favour that
model. The MC variant adds these rewritten mutants to the M-Only fitting pairs
while the test set stays fixed, so comparing the two shows how much the rewrites
change the result. Averaged over models, every measure we report changes little
when they are added (\Cref{sec:rq2-confounds});
since an average could hide a model-specific effect, \Cref{tab:confound-deltas} in
the appendix repeats the comparison per model: no model's accuracy moves by more
than $3.4$ points on any measure, and the probed Qwen model moves slightly
 \emph{against} the rewrites on every such measure. Thus, any family-specific
advantage is small. Incorrect programs also
differ in origin, pre-generated in RQ1 and mutations of the canonical solution in
RQ2; studying both exposes how this source shapes the direction.

\textbf{External Validity.} We study four 7--8B-parameter LLMs on two Python benchmarks, covering \indistribution and \outofdistribution stimuli. The observed effects may not hold for larger or smaller models, other architectures, or other programming languages, and our
single-function tasks do not cover repository-level settings such as
SWE-bench~\cite{jimenez2024swebench}. To mitigate this, we include both general
and code models, benchmarks of differing complexity, and \outofdistribution data.
Whether a single correctness direction generalizes across data is debated: some report a near-universal truthfulness hyperplane~\cite{liu2024universal}, others find uneven generalization across logical transformations~\cite{bao2025probing} or argue the property is graded and not a single boundary~\cite{ying2026truthfulness}. This is consistent with our finding that no single configuration is best across both benchmarks.

\section{Related Work}
\label{sec:related}

Prior work introduced the setting we build on: reading a model's hidden states to
judge the correctness of generated code without executing
it~\cite{llm-correctness-icse2026}. Bui et al.~\cite{bui2025openia} report that
internal representations correlate strongly with whether a program passes its
tests. Vu et al.~\cite{vu2026autoprobe} raise transferability by choosing the
read-out layer and token per model rather than fixing them, while others read the
same signal before generation to predict failure~\cite{lugoloobi2026encode}, as
a risk score over internal states~\cite{huang2025risk}, or as span-localized
calibrated uncertainty~\cite{gros2025localized}. Mechanistic accounts ask where
and how information about code correctness is represented inside the model, tracing it
through sparse autoencoders~\cite{tahimic2025mechanistic} and attribution
graphs~\cite{he2026codecircuit}, extending evidence that code models encode
program semantics internally~\cite{jin2023emergent}, while a parallel security
line reads or steers a vulnerability direction in the same
space~\cite{he2023sven, yu2025mixture, li2025steering, melo2025sparse}. Each of
these fixes one construction of the direction and inherits whatever correct and
incorrect programs its benchmark supplies; we instead measure how much the
read-out location, construction method, and prompt framing change the signal,
and find that the best location reverses between \he{} and \bcb{}. The closest
contrast, Vu et al.'s per-model selection~\cite{vu2026autoprobe}, sidesteps this
choice rather than quantifying its cost.

The probing recipe itself originates outside code, where hidden states linearly
separate true from false statements~\cite{azaria2023internal, marks2023geometry}
and representation engineering generalises it to reading and steering many
concepts~\cite{zou2023representation}. The same construction has since been
applied to a model's own outputs: probing reasoning traces for
self-verification~\cite{zhang2025reasoning}, predicting answer accuracy before
any answer is produced~\cite{cencerrado2025answer}, and reading correctness from
class centroids in the spirit of \meandiff~\cite{cho2026confidence}. We bring
this line to the correctness of generated code.

Our second question needs correct and incorrect programs that differ only in the
fault-causing change, which mutation testing supplies by seeding single localised
faults into working code~\cite{just2014mutants}. Behaviour-preserving rewrites
such as renaming and reformatting are the standard control for surface
form~\cite{wang2022recode, ramakrishnan2022semantic, jain2021contrastive}. We
combine both: mutations isolate the fault, behaviour-preserving rewrites vary surface form.
Kassem et al.~\cite{kassem2026automated} prompt an LLM to inject one
semantic error from a fixed taxonomy into \he\ and \bcb\ solutions, discard the
variants that still pass the tests, and release 1,217 labelled pairs. We
construct our own pairs for two reasons:
\begin{enumerate*}[label=\textbf{(\arabic*)}]
    \item LLM injection leaves residual label noise, which they acknowledge,
    whereas our deterministic operators act on a typed AST and so keep every
    mutation localised and type-compatible by construction;
    \item the MC variant needs a behaviour-preserving rewrite \emph{of each
    mutant} (\Cref{sec:rq2-data}), and no released dataset we know of pairs its
    faults with such rewrites.
\end{enumerate*}

\section{Conclusion}
\label{sec:conclusion}

We studied how the construction method, prompt framing, and hidden-state location
shape a captured code correctness direction (RQ1), and whether representing the
bug-causing change through controlled modifications beats
plausible failing attempts (RQ2). Two findings stand out: no single
configuration is best, with only \meandiff\ generally ahead of PCA; and isolating
the fault does not help, as a direction that tells a program from a
perturbed copy is not suited to judge an independent candidate.
Several directions follow from these results. First, RQ2 indicates that telling a
solution from a perturbed copy of itself is not the same as judging an independent
attempt; fault synthesis that better imitates the mistakes models make, e.g. via
semantics-aware or LLM-based mutation, could narrow that difference and make
controlled pairs a viable fitting source. Second, our study characterises the
captured direction only by its accuracy; a mechanistic account, e.g. through sparse
autoencoders, could reveal whether the direction encodes a semantic notion of
correctness or a bundle of correlated cues, and so explain why it shifts with the
configuration. Third, correctness is only one property of generated code; the same
probing methodology could target properties for which tests provide no executable
proxy, such as efficiency or security. Overall, reading correctness from a model's hidden states is possible, yet configuration-dependent.

\appendix[Per-Model Effect of the Confound Rewrites]
\label{sec:appendix-confounds}

\begin{table*}[!t]
  \centering
  \caption{Per-model view of the confound check: \val\ for a \meandiff\
  direction fit on a benchmark's controlled pairs without the rewrites
  (M-Only) and with them (MC), followed by the change in percentage points
  ($\Delta =$ MC $-$ M-Only). Settings follow \Cref{tab:rq2-transfer} (weighted
  sampling, averaged over read-out locations and framings), and the \textit{All}
  column reproduces its model-averaged cells. Two kinds of row support no
  conclusion and are marked accordingly: $\dag$ scores each variant on its own
  held-out pairs, so the two accuracies come from different test sets; $\ddag$
  leaves every model near the $25\%$ chance level of the 1-of-4 selection, so
  neither variant transfers there. The four unmarked rows share a test set and
  reach $60.7\%$, and carry the comparison.}
  \label{tab:confound-deltas}
  \footnotesize
  \setlength{\tabcolsep}{3pt}
  \begin{tabular}{@{}l rr rr rr rr rr@{}}
    \toprule
    & \multicolumn{2}{c}{Mistral} & \multicolumn{2}{c}{Qwen} & \multicolumn{2}{c}{CodeLlama} & \multicolumn{2}{c}{OpenCoder} & \multicolumn{2}{c}{All} \\
    \cmidrule(lr){2-3}\cmidrule(lr){4-5}\cmidrule(lr){6-7}\cmidrule(lr){8-9}\cmidrule(lr){10-11}
    tested on & M-Only & MC $(\Delta)$ & M-Only & MC $(\Delta)$ & M-Only & MC $(\Delta)$ & M-Only & MC $(\Delta)$ & M-Only & MC $(\Delta)$ \\
    \midrule
    \multicolumn{11}{@{}l}{\textit{fit on \bcb{} controlled pairs}}\\
    \bcb$^\ddag$               & $17.5$ & $17.4$ \dlt{-0.2} & $18.3$ & $18.2$ \dlt{-0.1} & $20.1$ & $21.4$ \dlt{+1.3} & $22.7$ & $19.3$ \dlt{-3.4} & $19.6$ & $19.0$ \dlt{-0.6} \\
    \quad layer on \bcb        & $33.7$ & $33.8$ \dlt{+0.1} & $32.4$ & $31.8$ \dlt{-0.7} & $33.7$ & $34.1$ \dlt{+0.5} & $31.7$ & $31.2$ \dlt{-0.5} & $32.9$ & $32.7$ \dlt{-0.2} \\
    \he                        & $26.2$ & $26.0$ \dlt{-0.2} & $38.6$ & $37.2$ \dlt{-1.4} & $25.2$ & $26.6$ \dlt{+1.4} & $29.3$ & $30.4$ \dlt{+1.1} & $29.8$ & $30.0$ \dlt{+0.2} \\
    held-out pairs$^\dag$      & $81.2$ & $79.8$ \dlt{-1.4} & $89.4$ & $87.9$ \dlt{-1.5} & $78.2$ & $75.2$ \dlt{-3.0} & $84.4$ & $84.1$ \dlt{-0.3} & $83.3$ & $81.7$ \dlt{-1.6} \\
    \addlinespace
    \multicolumn{11}{@{}l}{\textit{fit on \he{} controlled pairs}}\\
    \he                        & $39.3$ & $40.2$ \dlt{+0.9} & $60.7$ & $60.3$ \dlt{-0.4} & $37.8$ & $40.3$ \dlt{+2.5} & $52.5$ & $53.0$ \dlt{+0.5} & $47.6$ & $48.5$ \dlt{+0.9} \\
    \quad layer on \he         & $42.3$ & $43.0$ \dlt{+0.7} & $59.1$ & $58.2$ \dlt{-0.9} & $41.1$ & $40.9$ \dlt{-0.2} & $51.9$ & $53.9$ \dlt{+2.0} & $48.6$ & $49.0$ \dlt{+0.4} \\
    \bcb$^\ddag$               & $22.8$ & $21.2$ \dlt{-1.5} & $20.6$ & $19.1$ \dlt{-1.6} & $19.7$ & $19.3$ \dlt{-0.4} & $25.8$ & $25.8$ \dlt{+0.0} & $22.2$ & $21.3$ \dlt{-0.9} \\
    held-out pairs$^\dag$      & $63.0$ & $63.6$ \dlt{+0.6} & $84.2$ & $84.6$ \dlt{+0.4} & $64.1$ & $62.0$ \dlt{-2.2} & $75.0$ & $74.3$ \dlt{-0.7} & $71.6$ & $71.1$ \dlt{-0.5} \\
    \bottomrule
  \end{tabular}
\end{table*}

Recall that a confound is a behaviour-preserving rewrite of a \emph{mutant}, so it is itself
incorrect and fails in the same way as the mutant it comes from
(\Cref{sec:rq2-data}). Therefore, M-Only and MC are two fitting variants, the
second adding the refactored mutants to the pairs of the first. Each variant is
fitted, validated, and tested separately, so the M-Only and MC columns of
\Cref{tab:confound-deltas} hold the accuracies of two different directions, and
$\Delta$ is the difference between those two accuracies.

The M-Only/MC comparisons in \Cref{sec:rq2} are averaged over the four models.
The behaviour-preserving rewrites were generated by \texttt{Qwen2.5-Coder-32B-Instruct},
while the probed Qwen model in our experiments is \texttt{Qwen2.5-Coder-7B-Instruct}.
In principle, an average could hide an effect specific to the probed model,
so \Cref{tab:confound-deltas} shows both variants per model, for every measure
reported in \Cref{sec:rq2}.

Two things are visible in the absolute accuracies. First, every model
individually reproduces the pattern that \Cref{sec:rq2} reports on average: the
controlled pairs are highly separable in isolation. Yet, fitting on controlled pairs but testing on plausible failing attempts transfers poorly. Second, \texttt{Qwen2.5-Coder-7B-Instruct} is the most
separable model on five of the eight measures --- the two rows marked $\dag$ and
the three rows tested on \he\ --- but it already leads on them under M-Only, which
contains no rewrites at all. Therefore, its advantage is present before any
Qwen-generated data enters the fitting set, and so cannot be an effect of the
rewrites.

The deltas tell the same story. \Cref{tab:confound-deltas} marks the two kinds of row that support no conclusion, leaving four that do.

The two rows marked $\ddag$, both labelled \bcb, leave every model between
$17.5$ and $25.8\%$, around the $25\%$ that guessing gives on the 1-of-4
selection. When two variants
are that close to guessing, a small $\Delta$ only says that neither of them
transfers; it does not show that the rewrites are harmless. The two rows marked
$\dag$ have a different limitation: M-Only is tested on held-out M-Only pairs and
MC on held-out MC pairs, so their two accuracies come from different test sets
and cannot be compared directly. We report both kinds so that the table covers
every measure of \Cref{sec:rq2}.

The four unmarked rows carry the comparison: the two labelled \he\ and the two
labelled \emph{layer on}. In each of the four, M-Only and MC are scored on the same test set,
so the two differ only in the data used to build the direction. In the rows
labelled \he, that data is the variant's own controlled pairs: the layer is
selected on a held-out split of them, so M-Only and MC differ both in their
fitting pairs and in their validation pairs. In the rows labelled \emph{layer
on}, the layer is instead selected on a held-out split drawn from the test
benchmark itself, as RQ1 does for its in-distribution fits
(\Cref{sec:rq2-layer}). That split is the same whichever variant is being
fitted, so M-Only and MC are validated on identical data and their fitting pairs
are the only difference between them.
Accuracy across these four rows reaches $60.7\%$, well clear of guessing, so
there is a signal for the rewrites to disturb if they were going to. They do
not: adding them moves no model by more than $2.5$ points, and moves
\texttt{Qwen2.5-Coder-7B-Instruct} by at most $1.4$, always downwards.

Including the marked rows does not change the picture. Across all six rows that
test on a benchmark, no model gains or loses more than $3.4$ points, and
\texttt{Qwen2.5-Coder-7B-Instruct} is worse with the rewrites than without them
in every one of them, by at most $1.6$ points. Therefore, the per-model view 
agrees with the model-averaged result of \Cref{sec:rq2-confounds}: the rewrites
have little influence on any model's results, including the model whose family
produced them.

\balance
\bibliographystyle{IEEEtran}
\bibliography{refs}

\end{document}